\documentclass[preprint]{aastex}
\usepackage{amssymb}
\usepackage{stmaryrd}
\usepackage{amssymb,amsmath, epsfig, epsf, lscape}
\usepackage{bm}

\def\apj{ApJ}
\def\mnras{MNRAS}

\def\beq{\begin{equation}}
\def\eeq{\end{equation}}
\def\bey{\begin{eqnarray}}
\def\eey{\end{eqnarray}}

\def\mpc{\, h^{-1}{\rm {Mpc}}}

\def\kpc{\, h^{-1}{\rm {kpc}}}

\def\msun{\, h^{-1}{\rm M_\odot}}

\def\gs{\mathrel{\raise1.16pt\hbox{$>$}\kern-7.0pt
\lower3.06pt\hbox{{$\scriptstyle \sim$}}}}
\def\ls{\mathrel{\raise1.16pt\hbox{$<$}\kern-7.0pt
\lower3.06pt\hbox{{$\scriptstyle \sim$}}}}

\def\gtsima{$\; \buildrel > \over \sim \;$}
\def\ltsima{$\; \buildrel < \over \sim \;$}
\def\prosima{$\; \buildrel \propto \over \sim \;$}
\def\gsim{\lower.5ex\hbox{\gtsima}}
\def\lsim{\lower.5ex\hbox{\ltsima}}
\def\simgt{\lower.5ex\hbox{\gtsima}}
\def\simlt{\lower.5ex\hbox{\ltsima}}
\def\simpr{\lower.5ex\hbox{\prosima}}

\slugcomment{Submitted to {\it The Astrophysical Journal}}
\shorttitle{Flow pattern around dark matter halos} \shortauthors{Shi J.J. et al.}

\begin{document}
\title {Flow Patterns around Dark Matter Halos: the Link between
Halo Dynamical Properties and Large Scale Tidal Field}
\author{Jingjing Shi\altaffilmark{1,2}, Huiyuan Wang\altaffilmark{1} and H.J. Mo\altaffilmark{3}}
\email{jingssrs1989@gmail.com; whywang@mail.ustc.edu.cn; hjmo@astro.umass.edu}
\altaffiltext{1}{Key Laboratory for Research in Galaxies and Cosmology, Department of Astronomy, University of Science and Technology of China,
Hefei, Anhui 230026, China} \altaffiltext{2}{SISSA, Via Bonomea 265, I-34136 Trieste, Italy}\altaffiltext{3}{Department of
Astronomy, University of Massachusetts, Amherst MA 01003-9305,
USA}

\begin{abstract}
We study how halo intrinsic dynamical properties are linked to their formation
processes for halos in two mass ranges, $10^{12}-10^{12.5}\msun$ 
and $\ge 10^{13}\msun$,  and how both are correlated with the large scale 
tidal field within which the halos reside at present. 
Halo merger trees obtained from cosmological $N$-body simulations
are used to identify infall halos that are about to merge with their hosts.
We find that the tangential component of the infall velocity increases 
significantly with the strength of the local tidal field, but no strong correlation 
is found for the radial component. These results can be used to 
explain how the internal velocity anisotropy and spin
of halos depend on environment. The position vectors and velocities of infall
halos are aligned with the principal axes of the local tidal field, and the alignment
depends on the strength of the tidal field. Opposite accretion patterns are found in weak
and strong tidal fields, in the sense that in a weak field the accretion flow is dominated by radial
motion within the local structure, while a large tangential component is present in
a strong field. These findings can be used to understand
the strong alignments we find between the principal axes of the internal velocity
ellipsoids of halos and the local tidal field, and their dependence on the
strength of tidal field. They also explain why halo spin increases with
the strength of local tidal field, but only in weak tidal fields does
the spin-tidal field alignment follow the prediction of the
tidal torque theory. We discuss how our results may be used to understand the
spins of disk galaxies and velocity structures of elliptical galaxies and their
correlations with large-scale structure.
\end{abstract}


\section{Introduction}
\label{sec_intro}

In the standard cold dark matter (CDM) paradigm of structure formation,
a key concept is the formation and evolution of dark matter halos.
The halo properties, such as internal structures, dynamical properties,
assembly histories and clustering properties, and the correlations among them
have been studied in great detail \citep[see][for an overview]{Mo_etal2010}.
Since halos are the hosts of observed galaxies, these studies are essential
for understanding the formation and evolution of galaxies in the cosmic density field.

One particular property of the CDM halo population is that
the spatial clustering of halos of a given mass depends significantly
on various halo properties. \citet{Gao_etal2005} and \citet{LiMoGao2008}
found that old low-mass halos are more strongly clustered than their
younger counterparts. Halo clustering also depends on
halo structural properties, such as halo concentration and substructure
abundance \citep{Wechsler_etal2006, Jing_etal2007}, and on dynamical properties,
such as halo angular momentum \citep{Bett_etal2007, GaoWhite2007},
and internal velocity structure \citep{FaltenbacherWhite2010}.
All these dependencies, usually referred to as assembly bias, indicate
the importance of environmental effects on halo formation and evolution.

Attempts have been made to understand the environmental
effects on halo properties from various perspectives \citep{Wang_etal2007, Sandvik_etal2007, KeselmanNusser2007, Desjacques2008, Dalal_etal2008, FakhouriMa2009, LacernaPadilla2011, LacernaPadilla2012, Li_etal2013}. These
studies found that old low-mass halos usually reside in the vicinity of big structures,
and suggested that their accretion may be suppressed or even truncated
by the large scale tidal field. \citet{Wang_etal2011} studied the
correlations between a number of halo
properties and the tidal field, and found significant correlations
between the local tidal fields and all the halo properties they studied, including
half-mass assembly time, spin, axis ratio, and substructure abundance.
In particular, they found that the tidal field is the primary environmental
effect shaping most of the halo intrinsic properties,  while other
commonly used environmental indicators, such as the local mass density and
the morphology of the large scale structure, are secondary in that
their effects operate mainly through their correlations with the tidal field.
However, a detailed understanding of how environmental effects
shape the structure and dynamics of dark matter halos is still lacking.

In the CDM paradigm of structure formation, dark matter halos form
through the accretion (merger) of smaller halos, and halo properties
are expected to be determined by their formation histories. So far,
only a small number of simple quantities have been adopted to characterize
the formation histories of individual halos, and most of them are based
on characteristic times at which a halo has assembled a fixed fraction
of its final mass \citep[e.g.][]{Navarro_etal1995, Wechsler_etal2002,
Zhao_etal2009} or the gravitational potential well associated with the
halo has reached some depth \citep{Zhao_etal2003}. On the basis of
these formation times,  it has been shown that
younger halos on average are less concentrated and more elongated, spin
faster, and contain a larger amount of substructures, than their older counterparts
of the same mass \citep[e.g.][]{Gao_etal2004, Allgood_etal2006, Hahn_etal2007, Wang_etal2011}.
However, the formation histories of individual halos are complex and
cannot be described completely by these simple characteristic formation
times. Indeed, information about how small halos to be accreted (i.e. infall halos)
are distributed in phase space is totally lost in these characteristic
formation times, and yet may be pivotal in the understanding of the structural
and dynamical properties of the halos that grow through such accretion process.
Previous studies have found that infall halos on average have higher radial
than tangential velocities \citep{Tormen1997, Vitvitska_etal2002, Benson2005, Wang_etal2005, Wetzel2011, Jiang_etal2014}. Such anisotropic
orbits of accretion may affect  the internal velocity structure of the descendant
halos that form through such accretion. Indeed, dark matter halos in
$N$-body simulations are found to be dominated by radial orbits in their
internal velocity distributions, at least in the outer parts \citep{Colin_etal2000, Rasia_etal2004, Ludlow_etal2011, SparreHansen2012}.
Clearly, it is important to understand
and quantify such links between halo internal properties and their
accretion processes. Furthermore, since the phase space distribution of halos to be
accreted into a host halo is expected to be closely linked to the large-scale
environment within which  the host halo resides, such information is
also crucial in order to understand the environmental effects on halo
structure and dynamics.

In this paper, we study in detail how small halos are accreted by their hosts,
how the properties of the host halos are determined by the accretion process,
and how the accretion processes, through which the intrinsic properties of
halos are determined, are linked to the local environments of the halos.
We pay particular attention to halo dynamical properties, such as
velocity dispersion, angular momenta and velocity ellipsoid.
The structure of the paper is organized as follows. In Section \ref{sec_sim},
we describe the simulations we use, and our methods for halo identification,
merger tree construction and tidal field estimation.  In Section \ref{sec_eha},
we investigate the orbital distributions of infall halos and their dependencies
on the large scale tidal field. Section \ref{sec_edh} examines how
halo dynamical properties depend on environments, and how the dependence
can be understood in terms of the orbits of infall halos.
Finally, in Section \ref{sec_sum} we discuss and summarize our main results.

\section{Numerical Simulations and Dark Matter Halos}
\label{sec_sim}

\subsection{Simulation and Halo Identification}

In this study, we use two $N$-body cosmological simulations carried out
with Gadget-2 \citep{Springel2005}. These simulations adopted a flat
$\Lambda$CDM cosmological model, with $\Omega_{\Lambda,0}=0.742$ for the cosmological constant,
$\Omega_{\rm dm,0}=0.214$ and $\Omega_{\rm b,0}=0.044$
for CDM and baryons, respectively,  $h=0.72$ for the dimensionless
value of the Hubble constant, $\sigma_{8}=0.8$ for the {\it rms} linear mass
fluctuation in a sphere of radius $8\mpc$ extrapolated to $z=0$, and $n=1$
for the slope of the primordial fluctuation spectrum.
The CDM density field of each simulation is traced by
$1024^{3}$ particles, each with mass $m_{\rm p}\approx 5.3352\times10^{8}\msun$,
from $z=72$ to $z=0$ in a cubic box of a side length $200\mpc$.
The gravitational force is softened isotropically on a co-moving length
scale of $4\kpc$ (Plummer equivalent). Each simulation outputs 80
snapshots from $z=17$ to $z=0$, equally spaced in the logarithm
of the expansion factor.

Dark matter halos are identified using the standard friends-of-friends
(hereafter FOF) algorithm \citep{Davis_etal1985} with a link length
that is 0.2 times the mean inter-particle separation.
We only consider halos that contain at least 20 particles, and
the mass of a halo is the sum of the masses of all particles in the halo.
It is known that some FOF halos, in particular small ones containing
small number of particles, may be dominated by `fuzzy' particles
that are not gravitationally bound. We use SUBFIND algorithm
developed by \citet{Springel_etal2001} to identify subhalos that are
gravitationally bound. If the most massive bound structure contains less than
half of the total mass of the FOF halo, this FOF halo is considered
to be dominated by `fuzzy' particles and is excluded from our analysis.

\subsection{Merger Trees and Halos to be Accreted}\label{sec_sample}

Here we give a brief description of the construction of halo merger trees
and the identification of halos to be accreted into a host halo. We identify
dark matter halos using the method described above in each of the
snapshots and cross link halo particles in adjacent snapshots.  If more
than half of the particles in a halo (denoted as halo `A') end up in a halo
in the next snapshot (denoted as halo `B'), we call halo `A' a progenitor
of halo `B', and halo `B' the descendant of `A'.
This definition ensures that a halo can have one or more
progenitors but can only have one descendant. The uniqueness of the
descendant allows us to build up a unique merger tree for every halo at
present day. For any halo identified at a given time, its most massive
progenitor in the last snapshot is referred to as its main progenitor. Tracing
the main progenitors back in time gives the main trunk of the merger tree
of a halo identified at $z=0$.

For a given halo at present day, we select all of its progenitors
that are not parts of the main trunk but whose first generation
descendants are main trunk halos. These halos are
referred to as infall halos. The main progenitors, into which these
infall halos are falling, are called host halos. The redshift at which
an infall halo is identified is referred to as the infall redshift, $z_{\rm inf}$,
of the infall halo. Infall halos are therefore merging with their hosts
at a redshift around $z_{\rm inf}$. Some infall halos may have
been sub-halos of their hosts at $z>z_{\rm inf}$ but have
later moved outside of their hosts and are now falling back onto
the hosts. The orbits of these halos may have been severely altered by
interactions with the internal structures of the hosts, and so are
not suitable for our investigation of large-scale environmental effects.
Unfortunately, such halos cannot be directly identified because
our merger trees constructed by using FOF halos cannot trace the
evolution of subhalos within host halos. As an approximation, we adopt
the method developed in \citet{Wang_etal2009a} to identify these halos.
For an infall halo `A', we trace its main progenitor back in time until
its earliest main progenitor `B' is found in a snapshot, say $n$.
We then check whether or not more than half of the particles of halo
`B' belong to the main progenitor of the host halo of  `A'
at an earlier snapshot $n-1$. If yes, then `A' is considered to have
been ejected by the host at an early time,  and is excluded from
our analysis. Thus,  we only consider halos that are in their first infall.

A tiny fraction of infall halos are not contained in their host halos at $z=0$
\citep{Ludlow_etal2009, Wang_etal2009a, Bahe_etal2012, Li_etal2013}.
They are either ejected by their hosts after being accreted or
flybys fortuitously linked to their massive neighbors by the
FOF algorithm. This population may have important implications for
understanding the existence of quenched galaxies near clusters and
groups in the local Universe \citep{Wang_etal2009b, Li_etal2013, Wetzel_etal2014}.
To study the environmental dependence of halo assembly in detail, we identify
this population in the following way. If more than half of the particles in an infall
halo are not contained in its host (or descendant) at $z=0$, the infall halo
is thought to have finally escaped. This population will be referred to as ejected
halos. Other infall halos, which stay as subhalos within their hosts at $z=0$,
are referred to as the staying population.

Limited by finite mass resolution and small number statistics,
here we focus on the merger histories of host halos in two mass ranges,
$10^{12.5}\geq M_0\geq 10^{12}\msun$ (Milk Way size) and $M_0\geq 10^{13}\msun$ (massive group size),
where $M_0$ is the halo mass at $z=0$. The total numbers of host halos in
the two mass ranges are listed in Table \ref{tb1}. For the host halos in
the lower mass bin, infall halos with masses $M_{\rm inf}$ given by
$M_{\rm inf}/M_0 \geq 20m_{\rm p}/10^{12} \simeq 0.01$ are taken into account.
These infall halos are divided into two samples. The first sample,
denoted by M12(S), consists of only staying infall halos.
The second, M12(E), contains infall halos that eventually are
ejected by their hosts. The infall halos of the massive hosts are
divided  into four samples. The first, M13(S), consists of
all infall halos with $M_{\rm inf}/M_0\geq 0.01$ in the
staying population. The mass threshold adopted here is the same
as that for M12(S), and so one can investigate the dependence on
host halo mass by comparing M12(S) and M13(S). The second,
M13(S'), consists of infall halos with
$0.01>M_{\rm inf}/M_0\geq20m_{\rm p}/10^{13}\simeq 0.001$, again in
the staying population. A comparison between M13(S) and M13(S')
may help us to understand the dependence on the mass of infall halos.
The third and fourth samples, M13(E) and M13(E'), contain
infall halos of the ejected population,  with masses in the
same ranges as for M13(S) and M13(S'), respectively.
The numbers of infall halos in all the six samples are listed
in Table \ref{tb1}.

\begin{table*}
\caption[]{Number of host halos and infall halos in the samples. $M_0$ is the mass of host halos at $z=0$, in unit of $\msun$, $N_{h}$ is the number of host halos at $z=0$. $N_{\rm inf}$ is the number of infall halos. Low $z_{\rm inf}$ indicates $z_{\rm inf}\leq0.4$ and high $z_{\rm inf}$ indicates $0.75\leq z_{\rm inf}\leq1.25$.}
\smallskip
\begin{tabular}{ | c | c | c | c | c | c | c | c | c | c |}
\hline
$\log M_{0}$ & \multicolumn{6}{c |}{$\geq13$} & \multicolumn{3}{c |}{12 - 12.5} \\
\hline
$N_{h}$ & \multicolumn{6}{c |}{5793} &  \multicolumn{3}{c |}{37124}  \\
\hline
\hline
Infall halo &\multicolumn{2}{c |}{M13(S')} &\multicolumn{2}{c |}{M13(S)} & \multicolumn{1}{c |}{M13(E')} & M13(E) & \multicolumn{2}{c |}{M12(S)} & M12(E)\\
\hline
$N_{\rm inf}$ & \multicolumn{2}{c |}{214720} & \multicolumn{2}{c |}{49503} & \multicolumn{1}{c |}{37014} & 1718 & \multicolumn{2}{c |}{330221} & 30818 \\
\hline
\hline
$z_{\rm inf}$ & low & high & low & high & & & low & high & \\
\hline
$N_{\rm inf}$ & 54088 & 38251 & 12135 & 9772 & & & 56767 & 55946 &  \\
\hline
\end{tabular}
\label{tb1}
\end{table*}

\begin{figure*}
\centering
\epsfig{file=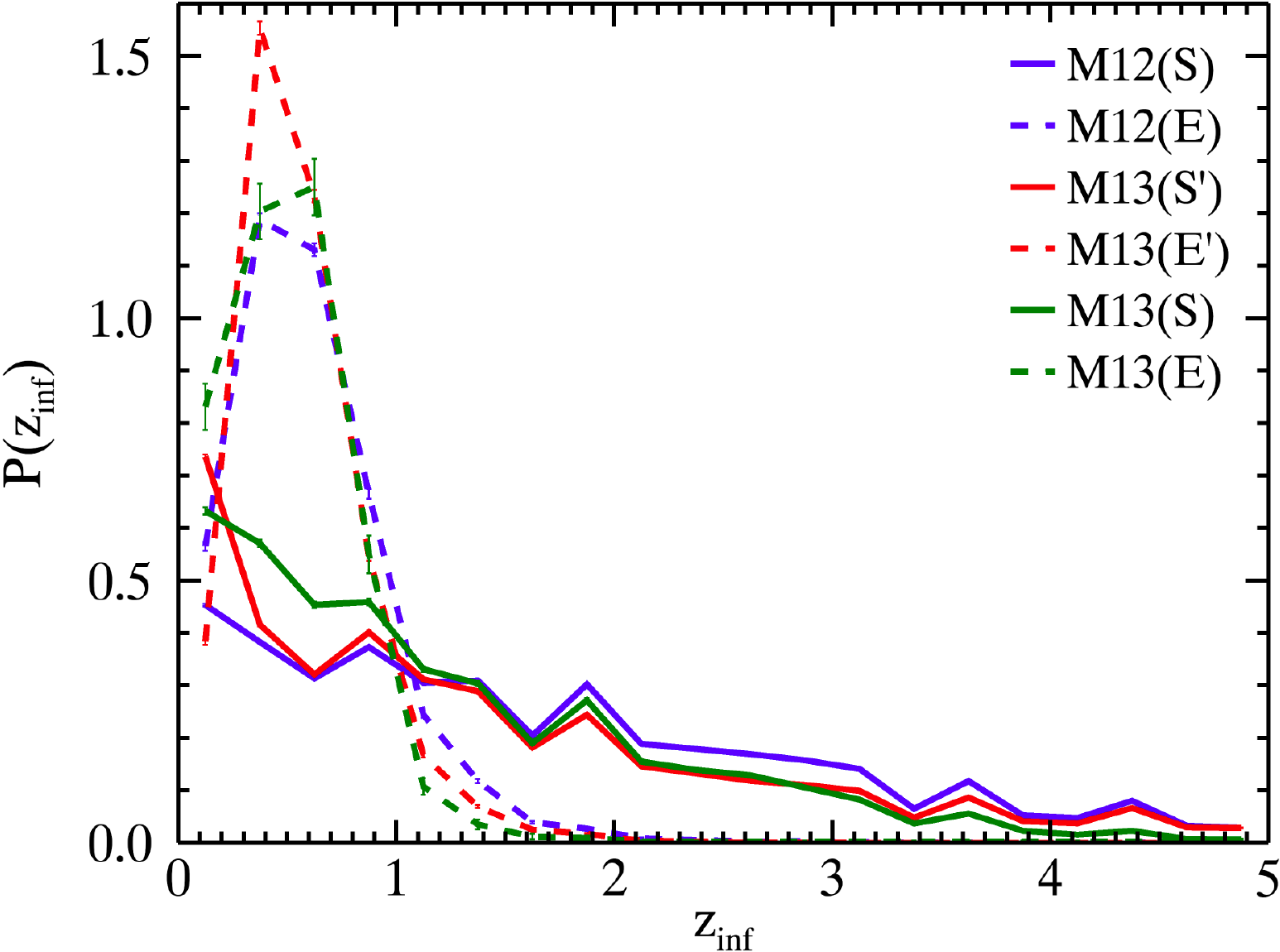,scale=0.8}
\caption{The probability distribution of infall redshift, $z_{\rm inf}$, for
the six infall halo samples as indicated in the figure.
See Section \ref{sec_sample} for sample selections.}
\label{fig_zinf}
\end{figure*}

Figure \ref{fig_zinf} shows the distributions of $z_{\rm inf}$ for the six
infall halo samples described above. M12(S) on average has
higher $z_{\rm inf}$ than M13(S), as expected from the fact that
smaller halos are, on average, older than more massive ones.
Infall halos that stay as sub-halos are accreted over
a wide range of redshift. In order to minimize possible dependence
on redshift, we consider infall halos accreted in two relatively
narrow redshift ranges, a low-redshift range $z_{\rm inf}\leq0.4$ and
a high redshift range, $0.75\leq z_{\rm inf}\leq1.25$.
The numbers of halos in these two redshift bins for samples M12(S),
M13(S) and M13(S') are also listed in Table \ref{tb1}.
The distributions of the ejected halos are much narrower,
with peaks at $z\sim0.5$,  and the majority of such halos
have infall redshifts below $z=1$. Because of this we do not split
ejected halos further according to infall redshifts.

\subsection{Large Scale Tidal Field}\label{sec_tdf}

A number of quantities can be used to characterize the large scale environment
of dark matter halos, including halo bias parameter, local mass over-density,
morphology of large scale structure (i.e. cluster, filament, sheet and void), velocity shear field
and large scale tidal field \citep[]{Mo_etal1996,Gao_etal2005,Maulbetsch_etal2007,Hahn_etal2007, Wang_etal2011, Libeskind_etal2013mn, Libeskind_etal2014a}.
In this paper, we adopt the (external) tidal field at the location
of a halo to represent the large-scale environment in which the halo resides.
The (external) tidal field is estimated by summing up the tidal forces exerting on
the halo by all other halos above a mass threshold, $M_{\rm th}=10^{12}\msun$,
and is normalized by the self-gravity of the halo in question \citep{Wang_etal2011}.
The local tidal field can be characterized by the three eigenvalues of the
local tidal tensor, $t_1$, $t_2$ and $t_3$ (by definition, $t_1>t_2>t_3$),
and the corresponding eigenvectors, ${\bm t}_{1}$, ${\bm t}_{2}$ and ${\bm t}_{3}$.
The three eigenvalues satisfy $t_1+t_2+t_3\equiv 0$, so $t_1$ is always positive
and $t_3$ is always negative. Thus, the large scale tidal field stretches the
material along ${\bm t}_{1}$ but compresses it along ${\bm t}_{3}$. In this paper,
we use $t_1$ as an indicator of the local tidal field strength.

The other method for calculating the tidal field, often adopted in the literature
\citep[e.g.][]{Hahn_etal2007, Zhang_etal2009}, directly makes use of
the mass density field to get the {\it mass tidal field}.
As shown in \citet{Wang_etal2011}, ${\bm t}_{1}$, ${\bm t}_{2}$
and ${\bm t}_{3}$ defined above are tightly aligned with the
corresponding eigenvectors of the mass tidal field.
Different from the mass tidal field, the tidal field defined above
does not include the contribution of the self-gravity of the halo,
and therefore is more closely related to the large-scale
environment. Moreover, two halos that reside in a similar
environment may suffer very differently from the local
environment. For example, a `hot'
environment for a small halo can be quite `cold' for a massive
halo. To take into account this halo mass-dependent effect,
our tidal field is normalized by the self-gravity of the halo,
so that one can compare the environmental effects for halos of
different masses. More details about the tidal field defined here
and its correlations with other environmental quantities can be
found in the appendix of \citet{Wang_etal2011}.

We consider halos accreted at different redshifts. 
The environmental indicator can be chosen to be either the tidal 
field within which the $z=0$ descendant halo resides 
or the tidal field when the accretion process occurs. 
In this paper, we use the tidal field at $z=0$ as an environmental 
indicator. There are two primary reasons for this choice. First, 
our eventual goal is to study similar effects in observational data 
(see Section \ref{sec_sum}). 
As shown in \citet{Yang_etal2007}, galaxy groups properly 
selected from large redshift surveys of galaxies can be used to 
represent the halo population. Dark matter halos are biased tracers 
of the underlying density field and can be used to estimate the 
large scale tidal field \citep{Wang_etal2012}.  
Currently, such a galaxy group catalog is only available 
at low redshift. Second, one of the purposes of this paper 
is to use the environmental dependence of halo accretion to 
interpret the environmental dependence of dynamical 
properties of halos at $z=0$ (Section \ref{sec_edh}).  
The local tidal field within which these halos reside 
provides one such environmental indicator that can be estimated 
from observation.

\begin{figure*}
\centering
\epsfig{file=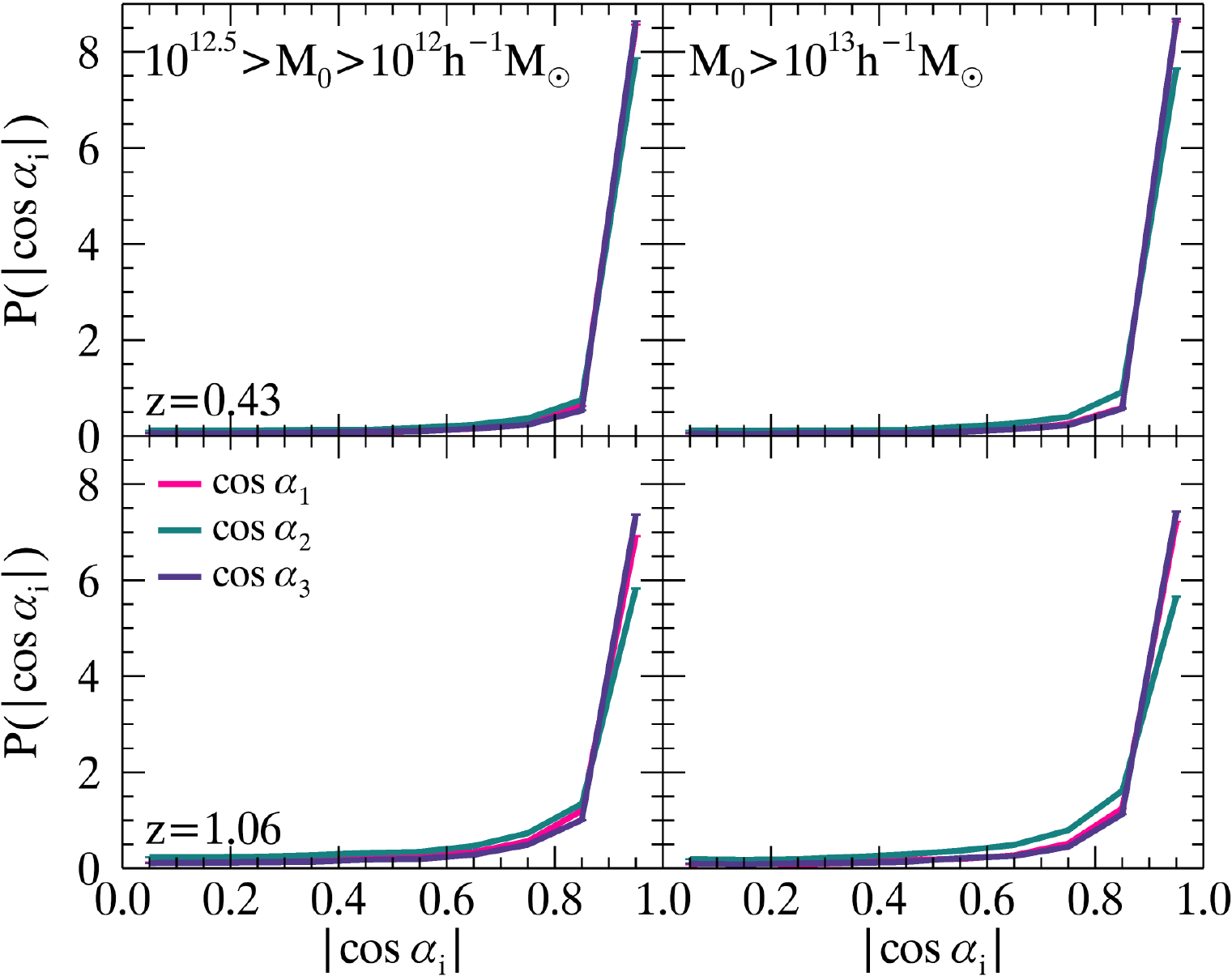,scale=0.8}
\caption{The probability distributions of $|\cos\alpha_i|$, where $\alpha_i$ is the angle between the 
eigenvectors ${\bm t}_{i} (i=1,2,3)$ of the tidal field around a $z=0$ halo 
and the corresponding eigenvectors around its main progenitors at $z=0.4$ (upper panels) and $z=1.0$ (lower panels). 
The right and left panels show the results for Milky Way sized and massive group sized halos, respectively.}
\label{fig_tii}
\end{figure*}

\begin{figure*}
\centering
\epsfig{file=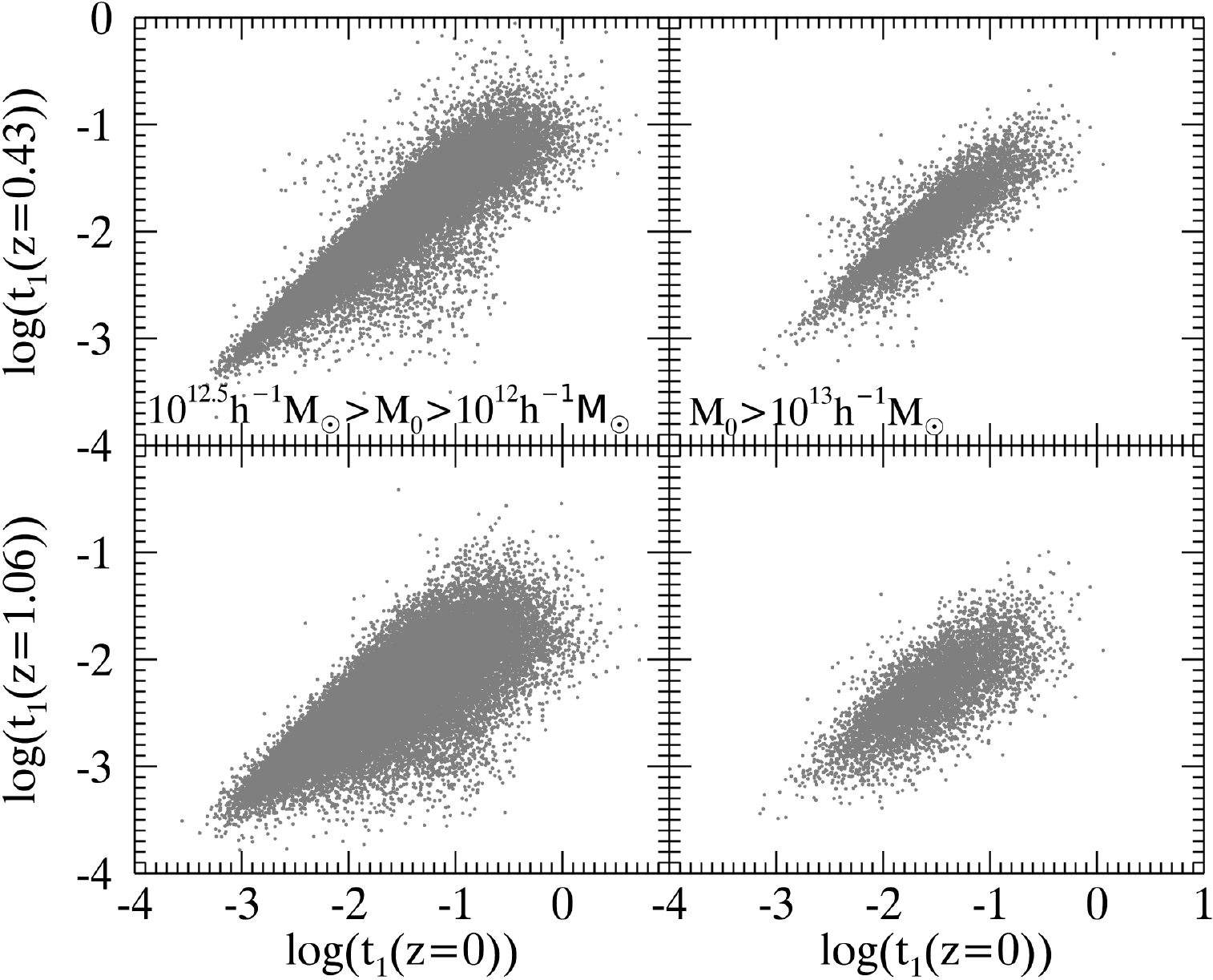,scale=0.8}
\caption{The comparison between the tidal field strength ($t_1$) around a $z=0$ halo and 
that around its main progenitors at $z=0.4$ (upper panels) and $z=1.0$ (lower panels). The right and left panels 
show the results for Milky Way sized and massive group sized halos, respectively.}
\label{fig_t1_comp}
\end{figure*}

One interesting question is how the tidal field around a 
halo evolves with redshifts. To answer this question, 
we analyze the alignments and correlations of the external 
tidal field around a $z=0$ halo with those around its main 
progenitors at $z=0.4$ and $z=1.0$. Note that the external tidal 
field at a high redshift is also calculated without including the 
contribution of surrounding halos that will end up in the final halo.
The results for the alignments and the correlations of $t_1$ 
are shown in Figure \ref{fig_tii} and \ref{fig_t1_comp}, respectively.  
The reason for choosing these two particular redshifts is that 
our following analyses focus on infall halos in the two redshift 
ranges, $z_{\rm inf} \leq 0.4$ and $0.75\leq z_{\rm inf}\leq1.25$. 
Clearly, the eigenvectors of the tidal fields at both $z=0.4$ and 
$z=1.0$ are strongly aligned with the corresponding vectors at 
$z=0$. The alignments between $z=0.4$ and $z=0$ are stronger than 
those between $z=1.0$ and $z=0$, and the dependence 
on halo mass is rather weak. 
For both halo mass ranges, the tidal field strength at $z=0.4$ 
is tightly correlated with that at $z=0$, and the correlation becomes 
weaker for $z=1.0$. Overall, the tidal field at $z=0$ can 
be used as a proxy of the tidal field at higher redshift, at least 
to $z\sim 1$, and particularly for the orientation of the tidal field.

\section{Environmental Dependence of Halo Accretion}
\label{sec_eha}

In this section, we investigate the environmental dependence of halo accretion
from three different aspects.
We emphasize again that we use the $z=0$ tidal field as our environmental indicator.
We first study the mass function of infall halos residing
in different environments (Subsection \ref{sec_mf}), and then investigate the correlations
between the tidal field and the orbital properties of infall halos (Subsection \ref{sec_orbit}).
Finally, in Subsection \ref{sec_ali}, we examine the alignment between the position and
velocity vectors of infall halos and the local tidal field.

\subsection{Infall Halo Mass Function}
\label{sec_mf}

\begin{figure*}
\centering
\epsfig{file=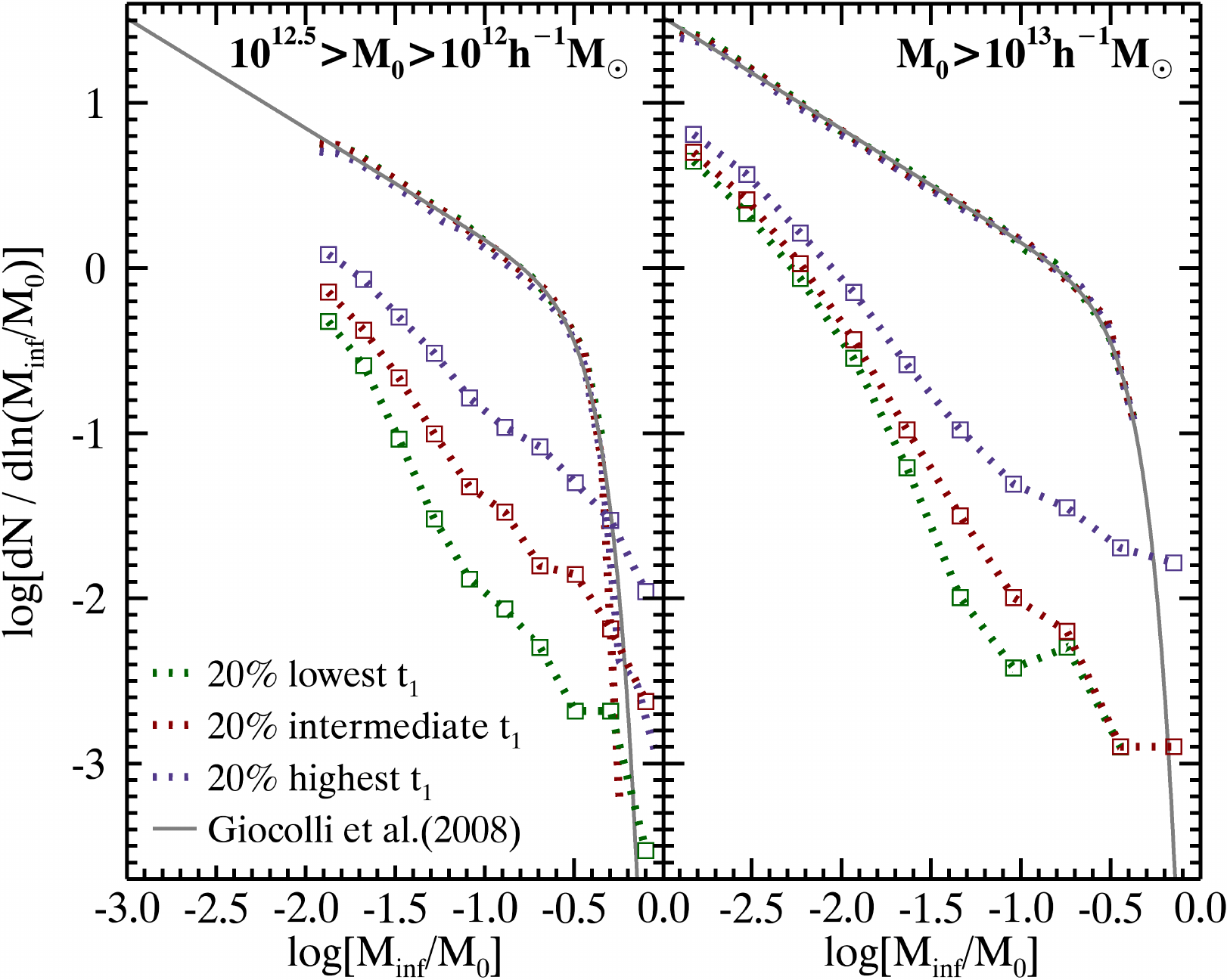,scale=0.8}
\caption{The comparison of infall halo mass functions in different environments. The blue, red and green lines correspond to the host halos in the lowest, intermediate, and highest 20 percentiles of
$t_{1}$. The left panel shows the results for host halos of $10^{12.5}\geq M_0\geq 10^{12}\msun$, and the right panel shows the results of $M_0\geq 10^{13}\msun$. The dotted lines show the staying
population of infall halos. Note that the results for the three $t_1$ samples almost
overlap. The dotted lines connecting squares are for ejected halos.
The gray lines are the best-fitting results based on the function given by
 \citet{Giocoli_etal2008}.}
\label{fig_mf}
\end{figure*}

The infall halo mass function, sometimes also called the un-evolved subhalo
mass function in the literature, is often used to study the evolution of subhalos
within their hosts \citep{Giocoli_etal2008, Yang_etal2011}.
Here we examine whether the infall halo mass function depends on the
large scale environment. To this end we calculate the mean infall mass
functions for host halos which are located in regions of the highest,
intermediate and lowest 20 percetiles of the $t_1$ distribution.
The results are shown in Figure \ref{fig_mf}. In each panel, the three
dotted lines show the mass functions of staying infall halos, while the dotted
lines connecting squares are the results for the ejected population.
To ensure completeness, we only use infall halos in samples M12(S)
and M12(E) to calculate the mass functions for host halos of
$10^{12.5}\geq M_0\geq 10^{12.0}\msun$, and samples
M13(S)$+$M13(S') and M13(E)$+$M13(E') for host halos of
$M_0\geq 10^{13}\msun$.

The mass functions of infall halos obtained
from samples M12(S) and M13(S)$+$M13(S') are almost
independent of $t_1$. Since these halos are the ones that will
stay in their hosts, they are the major sources of halo growth,
in the sense that the integration of the mass function should be
roughly equal to one. The mass function is also quite independent
of host halo mass, consistent with previous findings
\citep{Giocoli_etal2008}. We fit the simulation data with the
formula proposed by \citet{Giocoli_etal2008},
\begin{equation}
\frac{d N}{d\ln(m_{v}/M_{0})}=N_{0}x^{-\alpha}e^{-6.283x^{3}},
~~x=\frac{m_{v}}{\alpha M_{0}}
\end{equation}
where $m_{v}$ is set to be $M_{\rm inf}$.
The resultant mean fitting lines are shown in the figure
for comparison. As one can see, the empirical formula fits our
results well, demonstrating the robustness of our merger tree
construction. The mean fitting parameters are
$\alpha=0.67$ ($0.68$) and $N_{0}=0.43$ ($0.40$) for
M12(S) [M13(S)$+$M13(S')]. The slopes $\alpha$ obtained here
are slightly less than $\alpha=0.8$ obtained by \citet{Giocoli_etal2008},
but the amplitudes are significantly higher than their
value, $N_0=0.2$. The difference may be caused by different
cosmological models and the definition of halos in the
two analyses.

Different from the staying population, the mass functions for
M12(E) and M13(E)$+$M13(E') strongly depend on the large
scale tidal field. The ejected halo population is much more
abundant in regions of stronger tidal field, and the difference
becomes larger as the infall halo mass increases.
At the high mass end, the ejected halo abundance in the 20\%
highest $t_1$ regions is about 10 times higher than that at the
20\% lowest $t_1$ regions. This suggests that the
large-scale tidal field can affect the accretion of halos, and
infall halos in `hotter' (strong tide) environments are
more likely to escape from the potential well of their hosts.
The slope of the mass function for the ejected halo population
is, on average, steeper than the mass function of the staying
population, indicating that halos with lower masses are
easier to be ejected. Overall, the ejected population is
only a small fraction of the total, and the fraction is higher
for lower mass host halos (see Table \ref{tb1}).

\subsection{Orbits of Infall Halos}
\label{sec_orbit}

\begin{figure*}
\centering
\epsfig{file=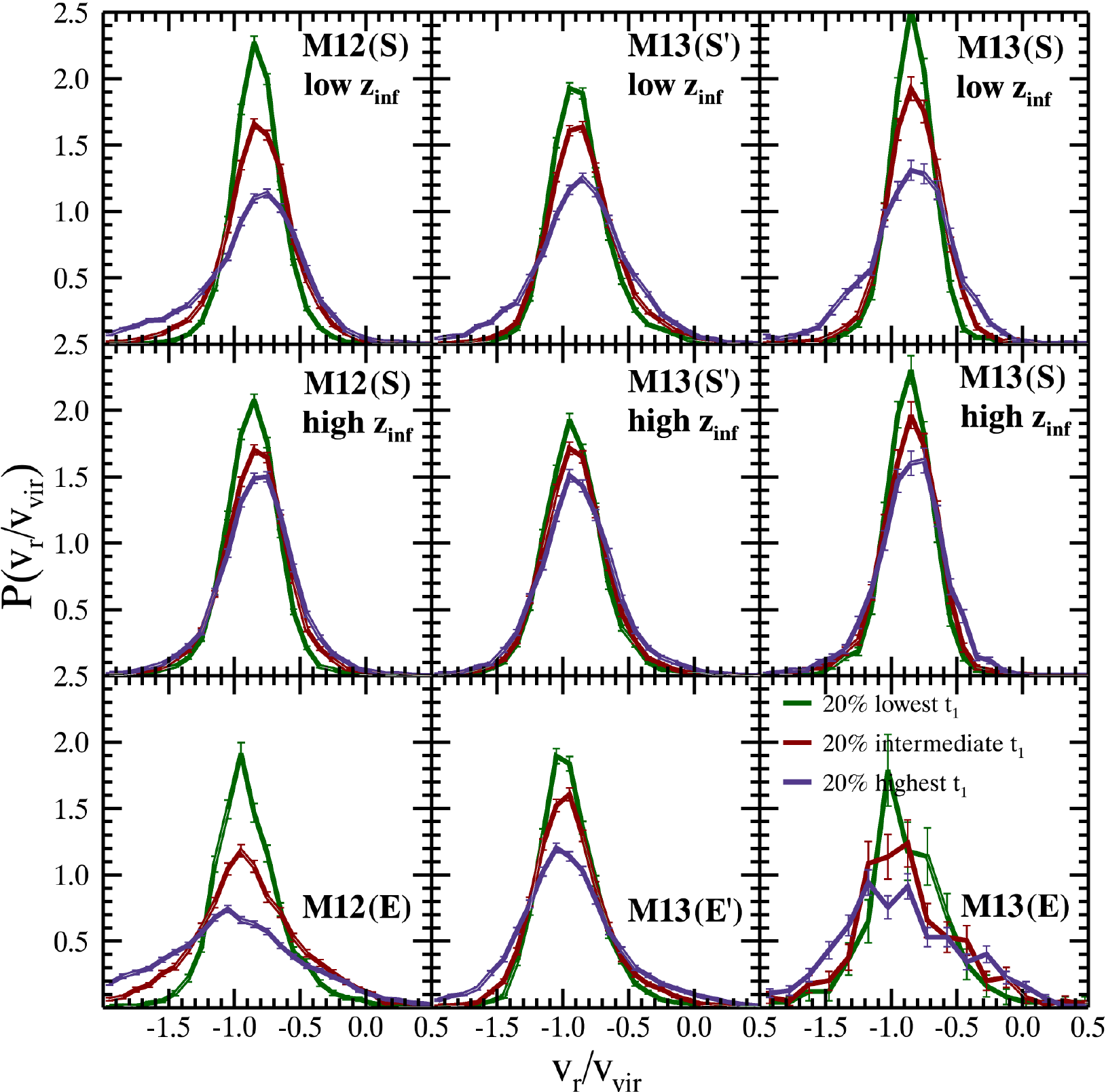,scale=0.8}
\caption{The radial velocity distributions of infall halos in different environments.
The blue, red and green lines show the results for host halos in the lowest,
intermediate and highest 20 percentiles of $t_{1}$.
Results are shown for infall halos in two different redshift ranges,
as indicated in each panel. The velocity is normalized by
$v_{\rm vir}$, the virial velocity of the host halo at the infall redshift.
The error bars are Poisson errors.
}
\label{fig_vr}
\end{figure*}

\begin{figure*}
\centering
\epsfig{file=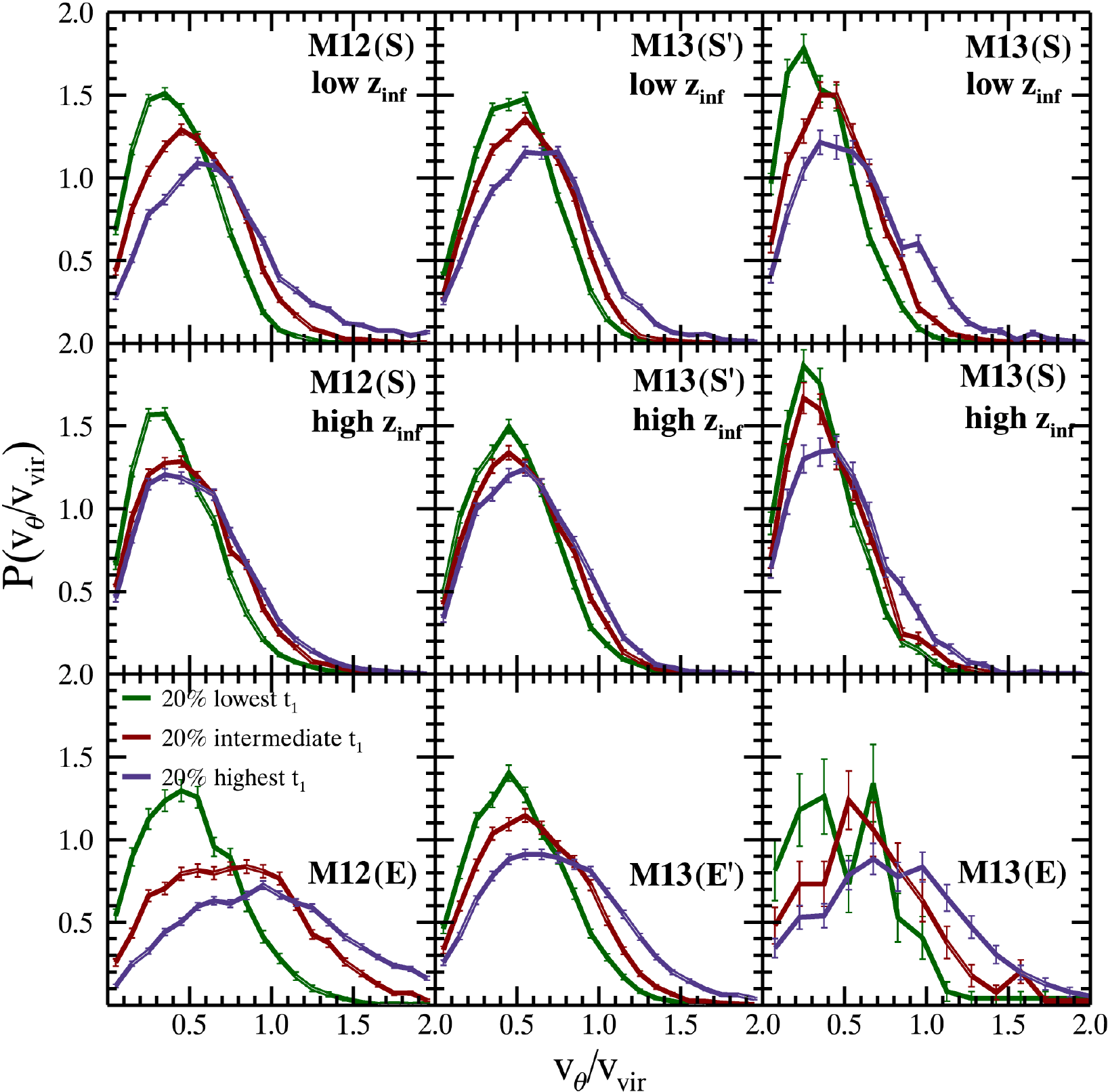,scale=0.8}
\caption{The same as Figure \ref{fig_vr} but for the tangential component of the infall velocity.}
\label{fig_vt}
\end{figure*}

Since the staying population dominates the total infall halos, we first
investigate how their acquisitions by their host halos are affected
by  environmental effects. Let us first look at $v_{\rm r}$ and
$v_{\theta}$, the radial and tangential velocities of infall halos
relative to the hosts at $z_{\rm inf}$. The radial direction is defined
as the position vector of the infall halo relative to the minimum potential
position of the host halo; a negative radial velocity means that the
halo is moving towards its host. Figure \ref{fig_vr} and \ref{fig_vt} show,
respectively, the probability distributions of $v_{\rm r}$ and $v_{\theta}$,
both normalized by the circular velocity of the host, $v_{\rm vir}$,
at $z_{\rm inf}$, for samples M12(S), M13(S) and M13(S') in the
highest, intermediate and lowest 20 percentiles of the $t_1$ distribution.
Results are shown separately for two narrow infall
redshift ranges, low ($z_{\rm inf}\leq0.4$) and high ($0.75\leq z_{\rm inf}\leq1.25$).
As one can see, the $v_{\rm r}$ distribution peaks around
$-0.9v_{\rm vir}$, while $v_{\theta}$ peaks at a smaller value.
These results are in qualitative agreement with those obtained
before \citep{Benson2005, Wang_etal2005, Wetzel2011, Jiang_etal2014}.
The fact that the free-fall velocity near the host virial radius
is about $-v_{\rm vir}$ suggests that the radial velocity of an infall
halo is primarily produced by the gravity of the host.
This interpretation is also supported by the similarity between
the distributions for host halos of different mass [samples M12(S) versus M13(S)]
at different redshifts. A small fraction of halos are moving
outward with very low (positive) velocities and are expected
to turn back shortly.

The gravity of the host is not the sole factor that affects the
velocity distributions of the infall halos. In fact, the distributions also
depend on the environments where the hosts reside. First, the
average tangential velocity increases as the tidal force increases.
As shown in the top left panel, the peak value of the tangential
velocity distribution for low redshift M12(S) increases
from $\sim0.3v_{\rm vir}$ for the lowest 20\% of $t_1$
to $\sim0.6v_{\rm vir}$ for the highest 20\%, in contrast to
the peak of the $v_{\rm r}$ distribution, which is almost independent
of $t_1$. Second, the distributions of both $v_{\rm r}$ and $v_\theta$
are broader in a stronger tidal field, and the effect is more significant
for radial velocity. Take the low-redshift M12(S) as an example, the
dispersion in the $v_{\rm r}$ distribution changes from
$0.19$ to $0.27$ and to $0.44$ from low $t_1$ to intermediate $t_1$ and to
high $t_1$, while the dispersion in the $v_{\theta}$ distribution
changes from $0.25$ to $0.30$ and to $0.32$.

The dependence on the tidal strength appears weaker for
infall halos at higher redshift. There are two possible reasons for
this. First, the tidal field, which is estimated from halos at $z=0$,
might not be a good tracer of environments at high redshift. Second,
environmental effects are indeed weaker at higher redshift.
We will come back to this question later.

\begin{figure*}
\centering
\epsfig{file=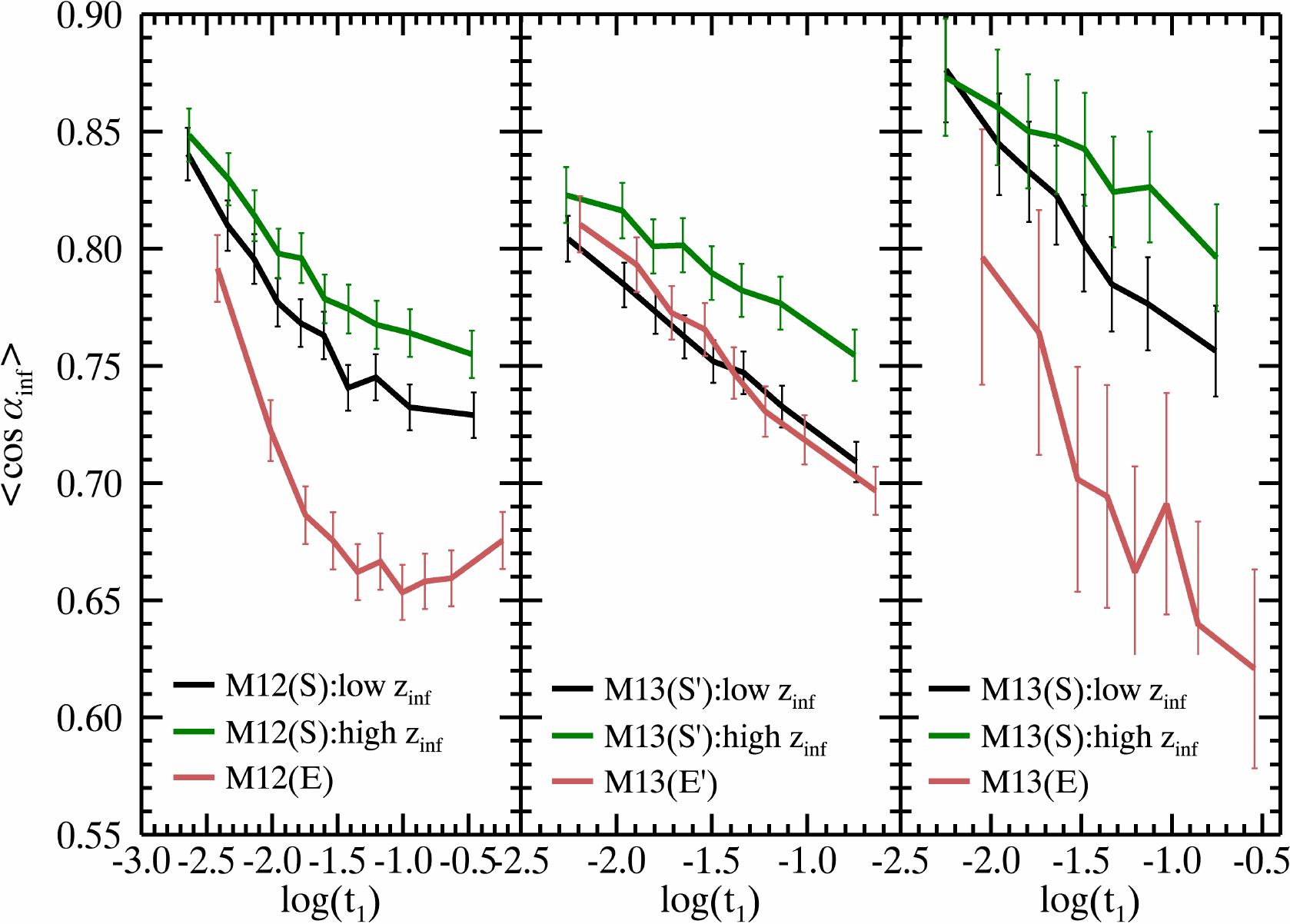,scale=0.8}
\caption{The mean cosine of the infall angle $\alpha_{\rm inf}$,
defined in equation (\ref{alpha_inf}),
as a function of the tidal field strength,
$t_1$, for different samples as indicated in the panels. For each curve,
the $t_1$ bin sizes are chosen so that each bin contains the
same number of halos. The error bars are Poisson errors.}
\label{fig_t1rv}
\end{figure*}

Another useful quantity is the infall angle, $\cos\alpha_{\rm inf}$, defined as
\begin{equation}\label{alpha_inf}
\cos\alpha_{\rm inf}=\frac{|v_{\rm r}|}{\sqrt{v^2_{\rm r}+v^2_{\theta}}}\,.
\end{equation}
We split the infall halo sample into several equal-sized subsamples
according to their local $t_1$, and calculate the mean values of
$\cos\alpha_{\rm inf}$ and $t_1$ for each of these subsamples.
The results are shown in Figure \ref{fig_t1rv}.
Note that radial velocity is the dominating component when
$v^2_{\rm r}\geq v^2_{\theta}/2$, i.e. when
$\cos\alpha_{\rm inf}\geq0.58$. The results clearly show that the
accretion flow preferentially moves radially in all environments.
Moreover, there is clear dependence of infall angle on the
redshift and on the masses of both the infall halo and the host.
The accretion flow at high redshift is more dominated by radial
motion than at low redshift. This is expected, because
environmental effects relative to the self gravity of the hosts
are weaker at higher redshift. A comparison between the results
for M12(S) and M13(S) suggests that the accretion flow around a
more massive host is also more radial.

The infall angle is strongly correlated with the strength of
the tidal force, $t_1$. As the tidal force increases, the mean
$\cos\alpha_{\rm inf}$ decreases significantly for all samples.
This is consistent with the velocity distributions shown
in Figures \ref{fig_vr} and \ref{fig_vt}. Interestingly, such environmental
dependence for infall halos accreted at high redshift ($z_{\rm inf}\sim1$)
is almost as strong as for those with lower $z_{\rm inf}$.
This suggests that the tidal field around a halo at $z=0$ is
correlated with the tidal field around its main progenitors at
high redshifts. As the large-scale structure in the Universe evolves,
the strength of the tidal field at the location of a halo is expected to
evolve with time. The tidal field obtained from the halo population
at $z=0$ may serve as an approximation of the \emph{scaled} version of the
tidal field at high $z$ (see Section \ref{sec_tdf} for more discussion).

Finally, let us look at the ejected halo population, whose results
are also presented in Figures \ref{fig_vr}, \ref{fig_vt} and \ref{fig_t1rv}.
Compared to the staying population, the ejected population has slightly
higher mean radial velocities, significantly higher tangential velocities
and much broader velocity distributions. These results are expected.
A higher tangential velocity means that the orbit is both
more loosely bound, which makes a final merger less likely,
and more circular, which makes orbital decay due to dynamical
friction less effective. Both effects make the sub-halo easier to
escape from the host. The dependence of
the tangential velocity distribution on the tidal field strength is
stronger for ejected halos than for the staying ones, as shown
in the lower panels of Figure \ref{fig_vt}, and the dominance
of the radial component of the infall velocity also decreases with
increasing $t_1$ more rapidly, as shown in Figure \ref{fig_t1rv}.

\subsection{Alignment of Accretion Flow with Tidal Tensor}
\label{sec_ali}

\begin{figure*}
\centering
\epsfig{file=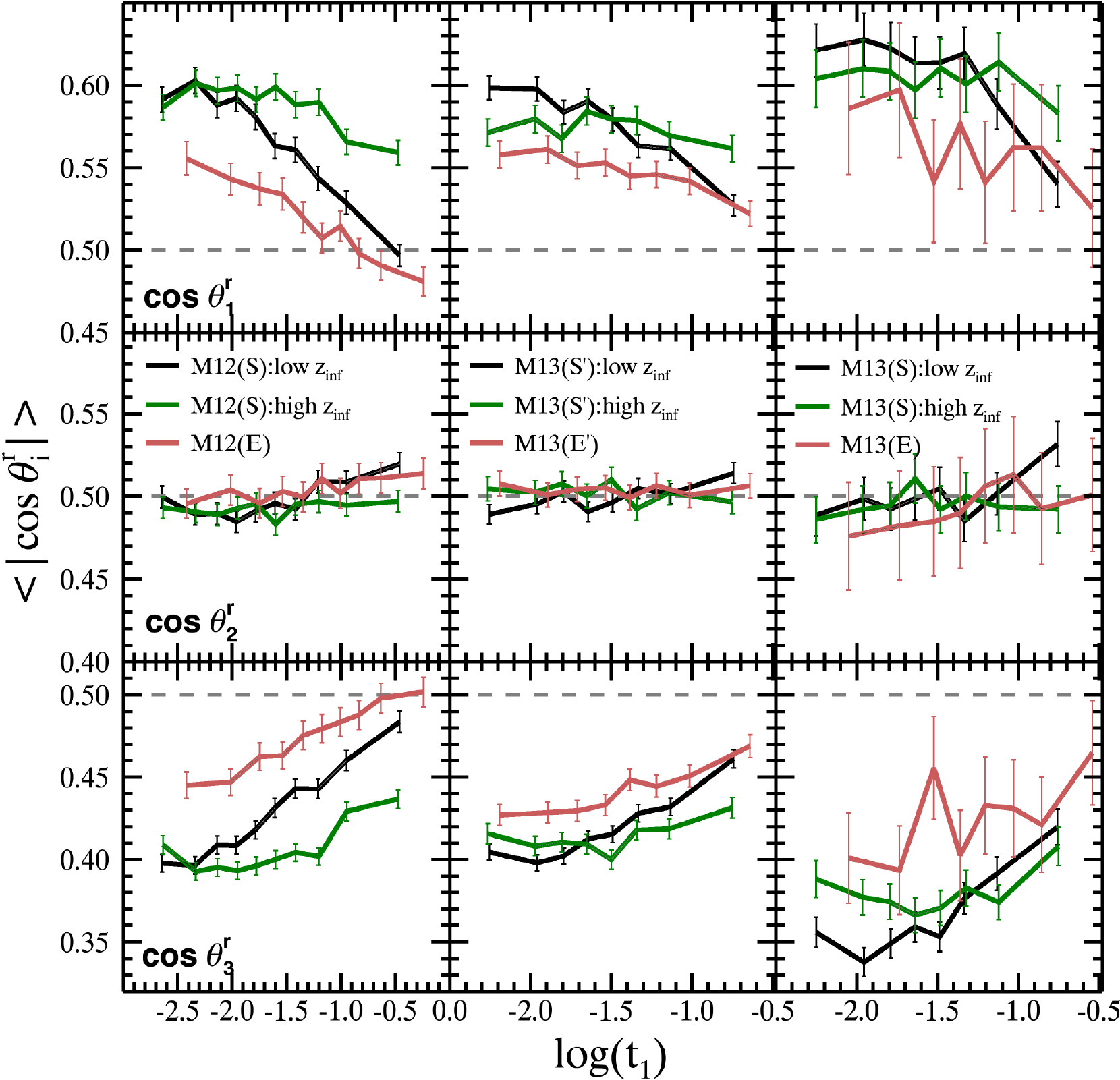,scale=0.8}
\caption{The mean of $\vert\cos \theta^{r}_{i}\vert$ ($i=1,2,3$)
defined in equation (\ref{theta_rvt}),
as a function of $t_1$.
From top to bottom, results are shown for $i=1$ (major axis),
$i=2$ (intermediate axis) and $i=3$ (minor axis). From left to right,
results are shown for different samples as indicated in the
intermediate-row panels. The dashed lines indicate isotropic distribution.
The error bars are Poisson errors.
}
\label{fig_t1rt}
\end{figure*}

The results shown above may suggest that the velocity field of accretion 
flow is regulated by the local large scale tidal field.\footnote{Note that correlation does not necessarily imply causation 
unless other possibilities are exhausted. A proof of causation, therefore, 
needs a much more detailed analysis.} Cosmological tidal field is known to be strongly anisotropic:
it stretches the accretion flow along the ${\bm t}_{1}$ direction, while
compresses it along ${\bm t}_{3}$. This suggests that the
spatial distribution and velocity field of infall halos are also
likely to be anisotropic, and perhaps have alignments with
the local tidal tensor. In this subsection, we investigate the
alignments of the three eigenvectors of local tidal tensor
with the position and velocity vectors of infall halos. The position
vector, ${\bm r}$, of an infall halo
is defined to be the vector from the minimum potential position
of its host to the infall halo itself, while the velocity vector, ${\bm v}$,
is defined to be its velocity relative to the velocity of the host. We use $\theta^r_i$
($\theta^v_i$) to denote the angle between the position (velocity)
vector and tidal eigenvector ${\bm t}_i$ ($i=1,2,3$), i.e.
\begin{equation}\label{theta_rvt}
\cos (\theta^r_i) = {{\bm r}\cdot {\bm t}_i \over
\vert {\bm r}\vert \vert {\bm t}_i\vert}\,;
~~~
\cos (\theta^v_i) = {{\bm v}\cdot {\bm t}_i \over
\vert {\bm v}\vert \vert {\bm t}_i\vert}\,.
\end{equation}
Here again, we first present the results for the staying population
of the infall halos. Figure \ref{fig_t1rt} shows the mean
$\vert\cos\theta^r_i\vert$ as a function of $t_1$.
It can be seen that the position vectors have a strong tendency to align
with ${\bm t}_1$ (the stretching direction) and to be perpendicular to
${\bm t}_3$ (the compressing direction), and are almost uncorrelated with
${\bm t}_2$. According to the definition of the tidal field, the large scale
mass distribution around a host halo tends to be in a filament
along ${\bm t}_1$ or within a sheet perpendicular to ${\bm t}_3$.
Thus, the accretion mass flows towards the hosts are expected
to be preferentially within these large scale structures,  and the
alignments shown in Figure \ref{fig_t1rt} follow directly from this
expectation. These results are consistent with that of \citet{Libeskind_etal2014b},
who found that mass accretion has the preference to be along the
direction of the weakest collapse, which is the ${\bm t}_1$ direction
defined here. Comparing the black and green lines in
Figure \ref{fig_t1rt}, we see that the average alignment signal is
stronger for infall halos at higher redshift, particularly for low-mass
hosts and for hosts located in high $t_1$ regions.
This is unexpected as the tidal tensor is estimated using
halos at $z=0$. On possible reason is that nonlinear effects,
which tend to suppress alignment, are weaker at higher $z$.
This interpretation is consistent with the fact that
the alignments are weaker in higher $t_1$ regions where nonlinear
effects are expected to be stronger. Regardless its origin, this
result suggests that our $z=0$ tidal field is a valid environmental
indicator for halos at high redshift (at least to $z\sim1$).

\begin{figure*}
\centering
\epsfig{file=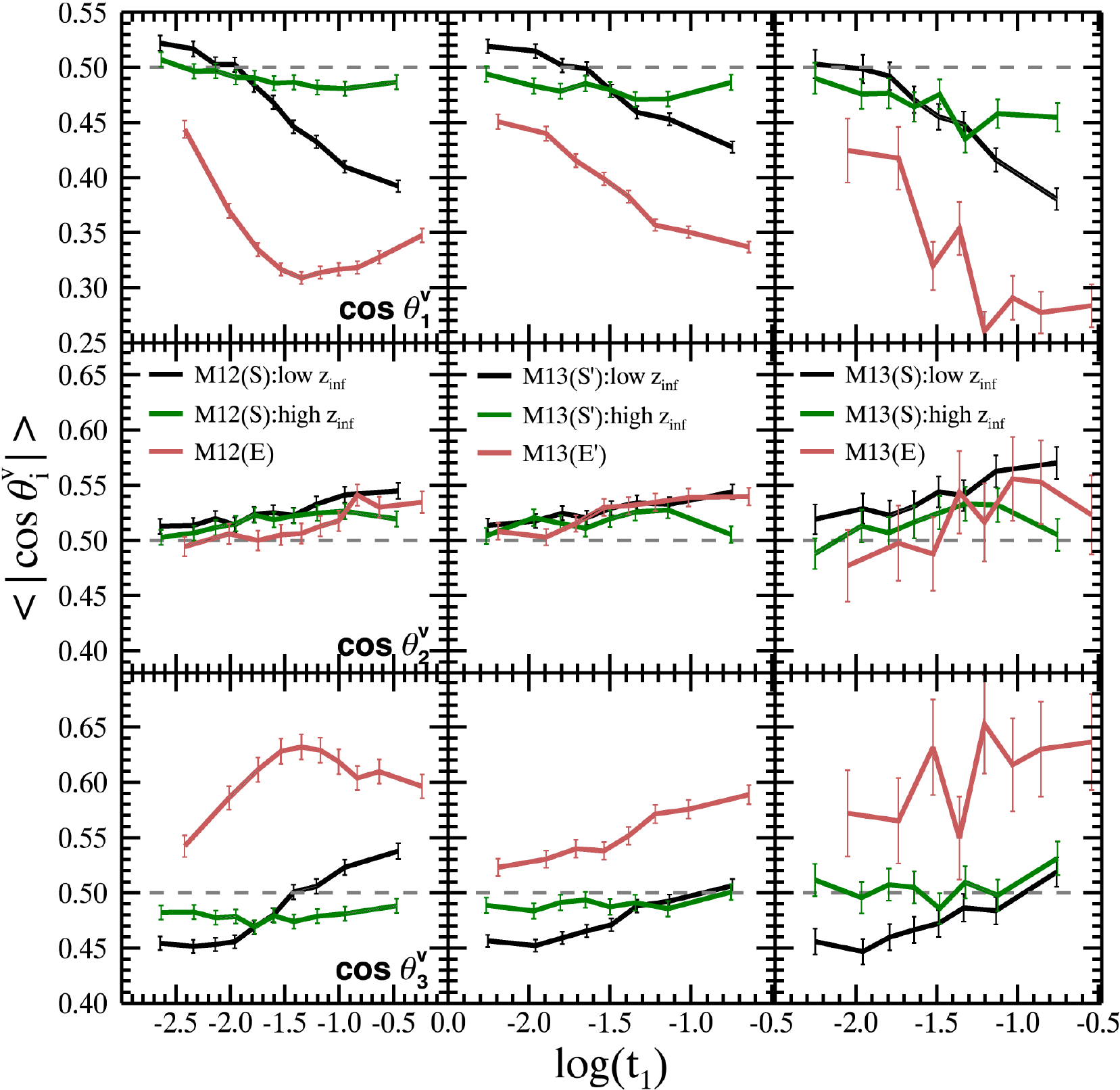,scale=0.8}
\caption{The same as Figure \ref{fig_t1rt} but for $\theta^{v}_{i}$,
defined in equation (\ref{theta_rvt}).
}
\label{fig_t1vt}
\end{figure*}

Figure \ref{fig_t1vt} show the mean $\vert\cos\theta_v^i\vert$ as a
function of $t_1$. Like $\vert\cos\theta^r_i\vert$ shown in
Figure \ref{fig_t1rt}, $\vert \cos\theta_v^i\vert$ shows
a strong correlation with the strength of the tidal force, and the
dependence is stronger for low-mass hosts at low redshift.
For example, for M12(S) at low redshift, there appears to be
two different accretion patterns depending on the environment
within which the host is embedded. In a weak tidal field, infall halos are
preferentially accreted along the directions that are parallel
with ${\bm t}_1$ and perpendicular to ${\bm t}_3$
(see the black solid lines in the left panels of Figure \ref{fig_t1rt}).
The velocity vectors of these infall halos have a weak
tendency to be parallel with ${\bm t}_1$ but a strong tendency
to be perpendicular to ${\bm t}_3$, as shown by
the black solid lines in the left panels of Figure \ref{fig_t1vt}.
In contrast, in a strong tidal field, infall halos are accreted
along directions that are almost uncorrelated with the
tidal tensor, while the velocity vectors tend to be perpendicular to ${\bm t}_1$
and parallel with ${\bm t}_3$. For sample M13(S) and M13(S'),
and for infall halos with high $z_{\rm inf}$, the overall trends
are very similar, albeit weaker.

\begin{figure*}
\centering
\epsfig{file=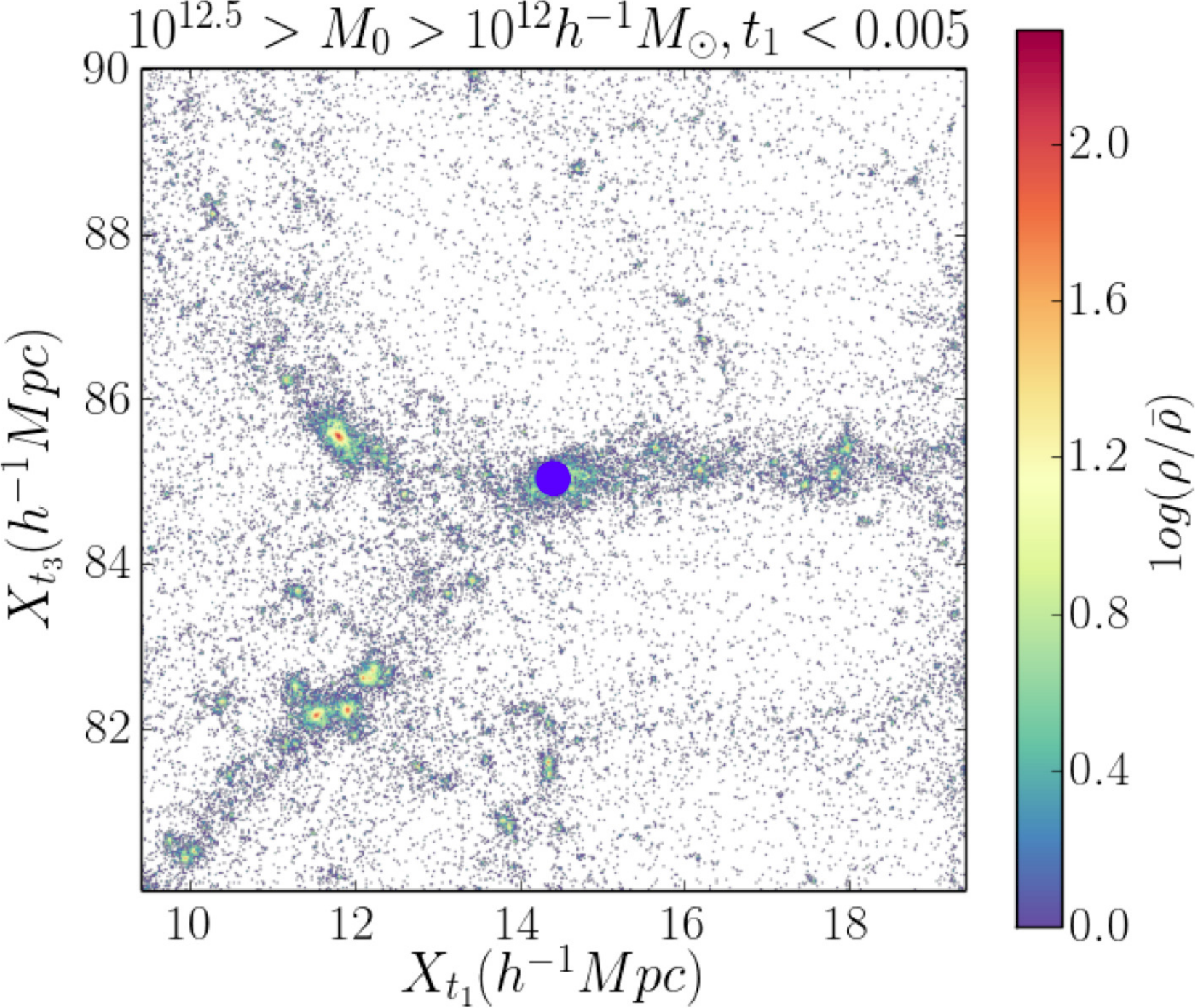,scale=0.45}
\epsfig{file=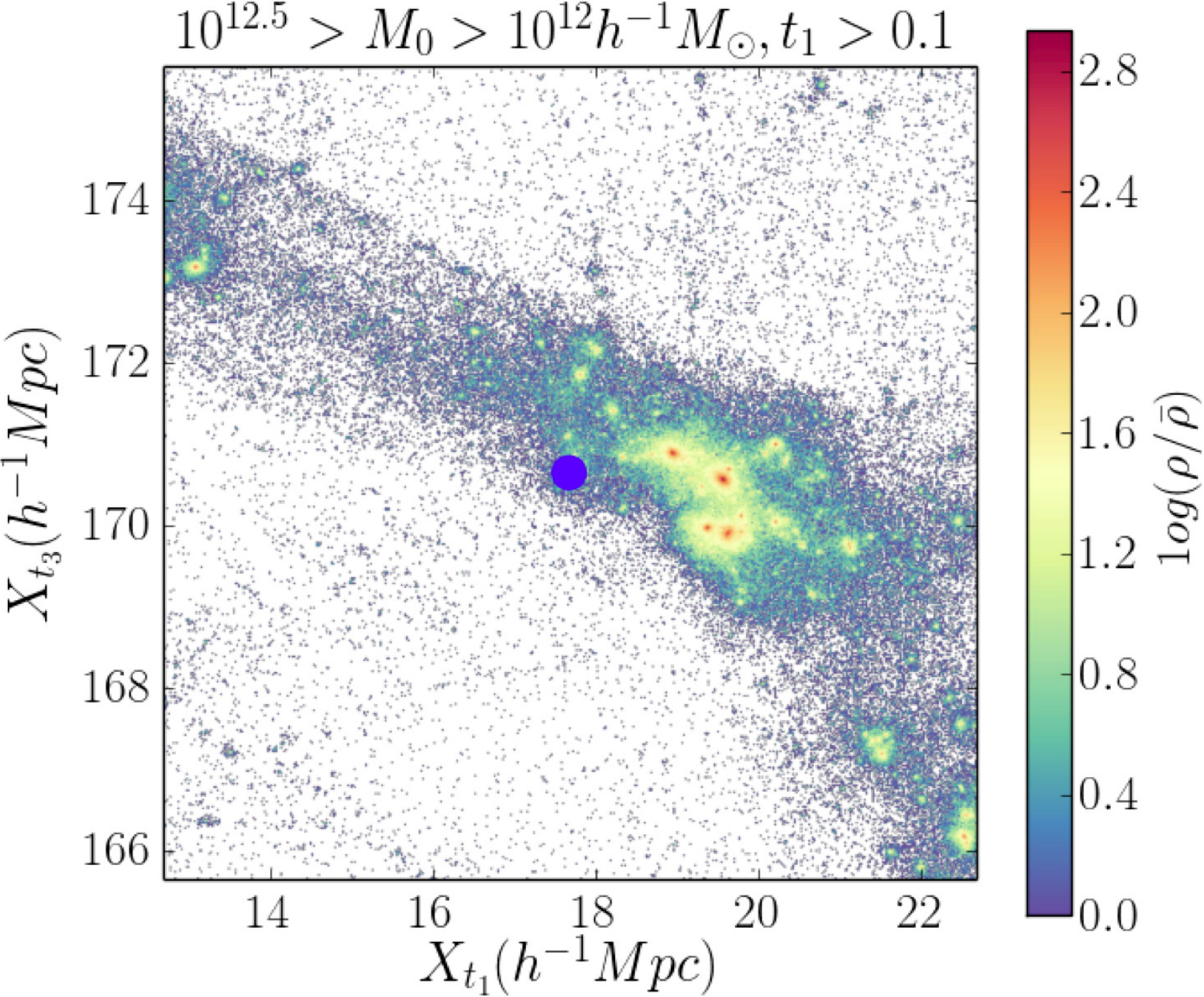,scale=0.45}
\caption{The projected density contrast maps around two Milk Way sized halos in weak and strong tidal fields 
as indicated in the panels. 
The two halos are shown as blue solid circles at the centers of the panels. 
The maps are shown in the ${\bm t}_1$- ${\bm t}_3$ plane, and 
the thickness is $5{h^{-1}\rm Mpc}$. The values of the density contrast are color coded, as shown in the color bars.   
}
\label{fig_lss}
\end{figure*}

The dependence on the tidal strength described above is interesting. In particular, why are the alignment signals
stronger for host halos located in weak tidal field where
large scale tidal field is expected to have a weak impact? 
The large scale (usually filamentary) structure 
surrounding a halo in a weak tidal field is not expected to be 
massive in comparison to the halo itself, and the thickness of 
the filamentary structure is likely to be comparable to the size of 
the halo (see Figure \ref{fig_lss} for an example).
For a host halo residing in a small filament, where the 
eigenvectors ${\bm t}_1$ and ${\bm t}_3$ are
expected to be parallel with and perpendicular to the filament,
respectively, the gravitational field
is dominated by the host halo itself. The infall halos, which
are located in the filament, are expected to have the tendency to
move along the filament as they fall onto the host, so that the
position and velocity vectors of the infall halos both have
the tendency to be aligned with the filament.
In this case, the role of the tidal field is to produce a `cold' filamentary 
structure from which the halo accrete new material. In contrast, for halo located in a strong tidal field,  
the surrounding structure is usually larger than the halo, 
even if it is a filamentary structure on a larger scale (see Figure \ref{fig_lss}). 
In this case,  the halo can accrete infall halos from different directions, 
producing a weak alignment between the position vectors and 
the tidal eigenvectors. In such an environment, the large scale tidal field
plays an important role in determining the motions of infall halos,
which generates deceleration of accretion along ${\bm t}_1$
and acceleration of accretion along ${\bm t}_3$. This explains
why in a strong tidal field, the velocities of the
infall halos tend to be parallel with ${\bm t}_3$ and perpendicular
to ${\bm t}_1$ (Fig. \ref{fig_t1vt}),
even though they still have a (weak) tendency to be
distributed along ${\bm t}_1$ at the time of accretion (Fig. \ref{fig_t1rt}).

The ejected halos exhibit similar dependence
of $\vert \cos\theta^r_i\vert $ on $t_1$. The mean alignment signal is
weaker than that for the staying population, indicating that the ejected
halos fall onto their hosts in a more isotropic manner.
The difference between ejected and staying populations is
particularly large in the velocities of infall halos, with the ejected
population showing a much stronger tendency of their velocity vectors
to be perpendicular to ${\bm t}_1$ and parallel with ${\bm t}_3$
in all environments. In the weak tidal field, this velocity - tidal
field alignment for ejected halos is opposite to that for the staying
population (Figure \ref{fig_t1vt}). These results together suggest
that ejected halos are a special population of infall halos
even before they are accreted by their hosts.

\section{Environmental Dependence of Halo Dynamical Properties}
\label{sec_edh}

In the previous section we have shown that the accretion patterns
of halos are correlated with the tidal field at $z=0$. Since the intrinsic
properties of dark matter halos are expected to depend on their
formation histories, halo intrinsic properties are expected to be
correlated with environment as well. \citet{Wang_etal2011} have investigated
the correlation between various halo structural properties with environment.
Here we focus on the dynamical properties of dark matter halos, such as
halo velocity structure and spin. There have been investigations about
how the velocity anisotropy, spin and velocity ellipsoid of halos
are affected by environment \citep[e.g.][FW10]{Bett_etal2007,GaoWhite2007,Hahn_etal2007,
Zhang_etal2009}. Our approach is different from
these studies in that we use tidal field as an environmental indicator.
More importantly, we try to interpret the environmental effect in
terms of the environmental dependence of the accretion we obtained above.

\begin{figure*}
\centering
\epsfig{file=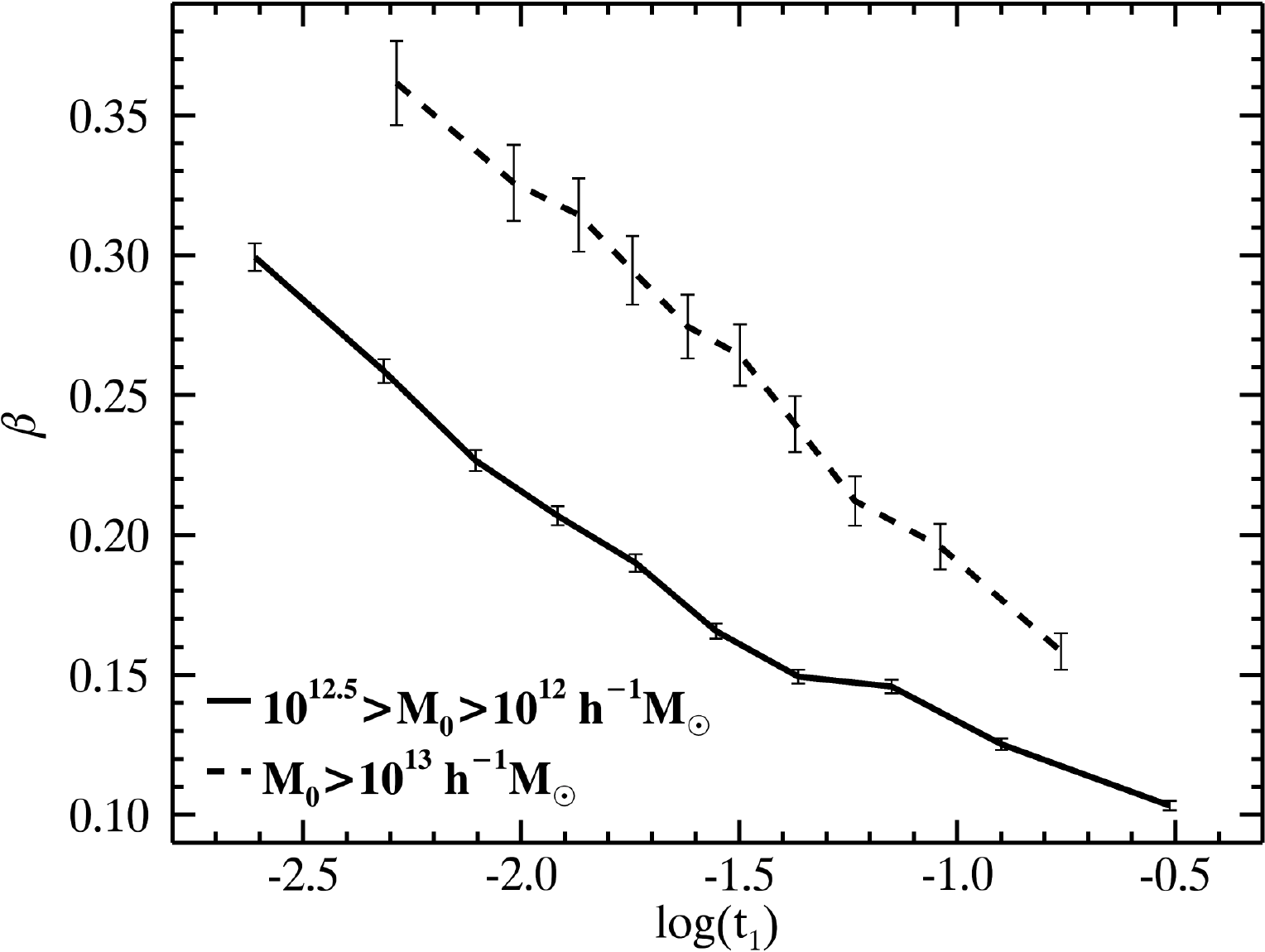,scale=0.8}
\caption{The correlation between velocity anisotropy, $\beta$ defined in equation (\ref{beta_def}),
of host halos and $t_1$ for two mass ranges. All $t_1$ bins
contain equal number of halos.
The curves connect the median values of
 $\beta$, while the error bars are Poisson errors.}
\label{fig_t1beta}
\end{figure*}

We first investigate the velocity anisotropy parameter, which is defined to be
\begin{equation}\label{beta_def}
\beta=1-{\sigma_{\theta}^{2}\over 2\sigma_{\rm r}^2}\,,
\end{equation}
where $\sigma_{\rm r}$ and $\sigma_{\theta}$ are, respectively,
the radial and tangential velocity dispersion, evaluated using
all halo particles. By definition, a negative (positive) value of
$\beta$ implies dominance of tangential (radial) motion, and
$\beta=0$ indicates an isotropic velocity field. Figure \ref{fig_t1beta}
shows $\beta$ as a function of $t_1$ for host halos at $z=0$
in two mass ranges. There is a clear trend that $\beta$ decreases
monotonically with increasing $t_1$. The internal velocity fields are
dominated by radial motion for halos in weak tidal fields,
and are almost isotropic for halos in strong tidal fields.
FW10 found that halos of low $\beta$ are more clustered than
those of high $\beta$. Given that the tidal field is on average
stronger in higher density regions \citep{Wang_etal2011},
our results are consistent with theirs. Note that for a given $t_1$,
$\beta$ is higher (meaning radial velocities is more dominating)
for higher mass halos.

\begin{figure*}
\centering
\epsfig{file=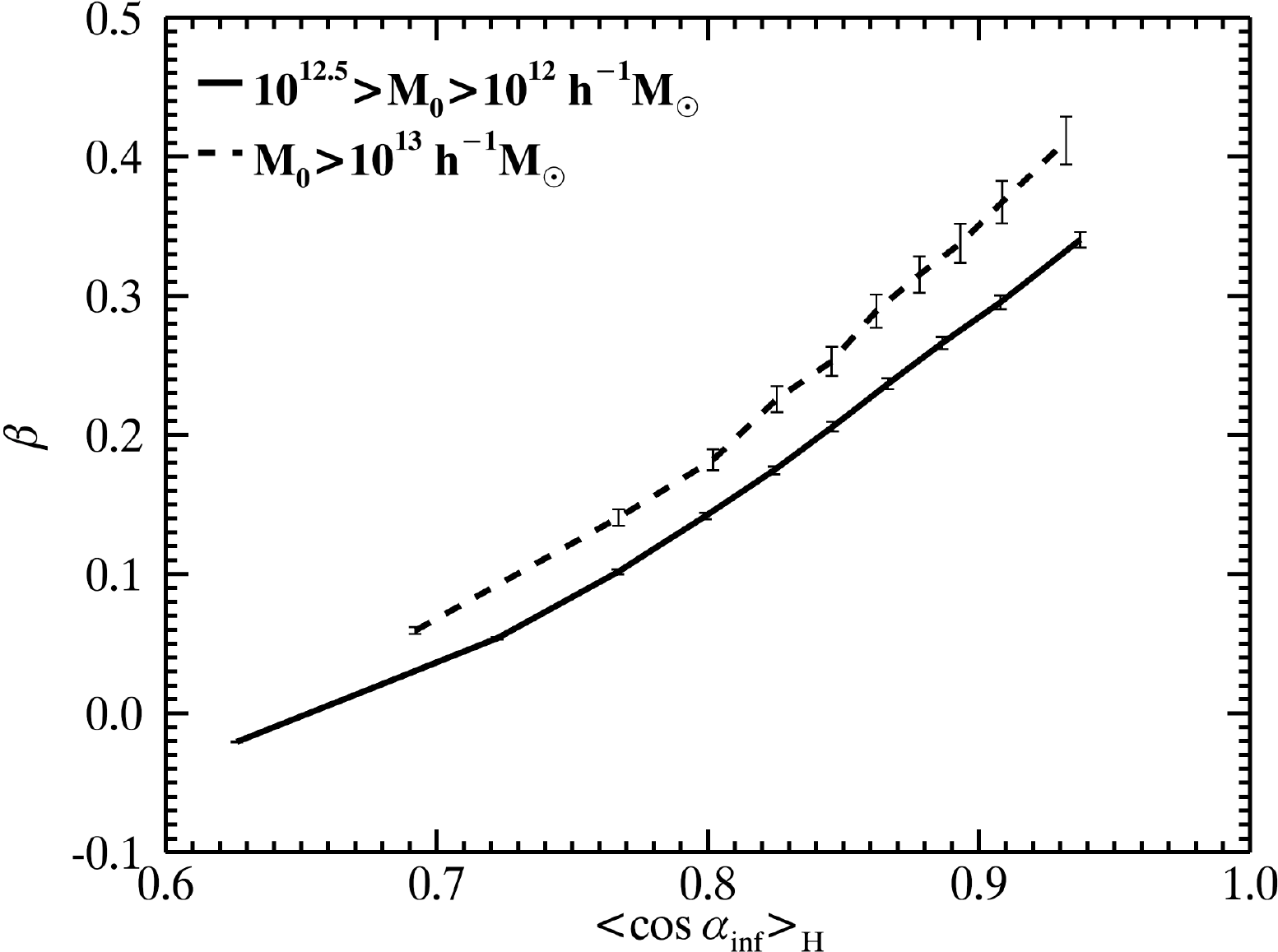,scale=0.8}
\caption{The correlation between the velocity anisotropy $\beta$ and
$\langle\cos\alpha_{\rm inf}\rangle_{\rm H}$.
For each host halo, $\langle\cos\alpha_{\rm inf}\rangle_{\rm H}$ is the
mass-weighted average value over all of its infall halos in the staying
population. All $\langle\cos\alpha_{\rm inf}\rangle_{\rm H}$ bins
contain equal number of halos.
The curves connect the median values of
 $\beta$, while the error bars are Poisson errors.}
\label{fig_rvb}
\end{figure*}

The velocity anisotropy of halos very likely reflects the anisotropy
in the velocity distribution of infall halos. As a test of this,
we calculate the mean cosine of the infall angle,
$\langle\cos\alpha_{\rm inf}\rangle_{\rm H}$, weighted
by the infall halo mass, for each host halo and show
$\beta$ versus $\langle\cos\alpha_{\rm inf}\rangle_{\rm H}$
in Figure \ref{fig_rvb}. Here, only staying infall halos are used
to calculate $\langle\cos\alpha_{\rm inf}\rangle_{\rm H}$.
We see a very strong positive correlation between these
two quantities. The more tangential the mean orbit of
the infall halos is, the more dominated the host halo is by
tangential motions. Given the strong correlations between
$\cos\alpha_{\rm inf}$ and $t_1$ shown in Figure \ref{fig_t1rv},
the dependence of $\beta$ on $t_1$ is straightforward
to understand.

\begin{figure*}
\centering
\epsfig{file=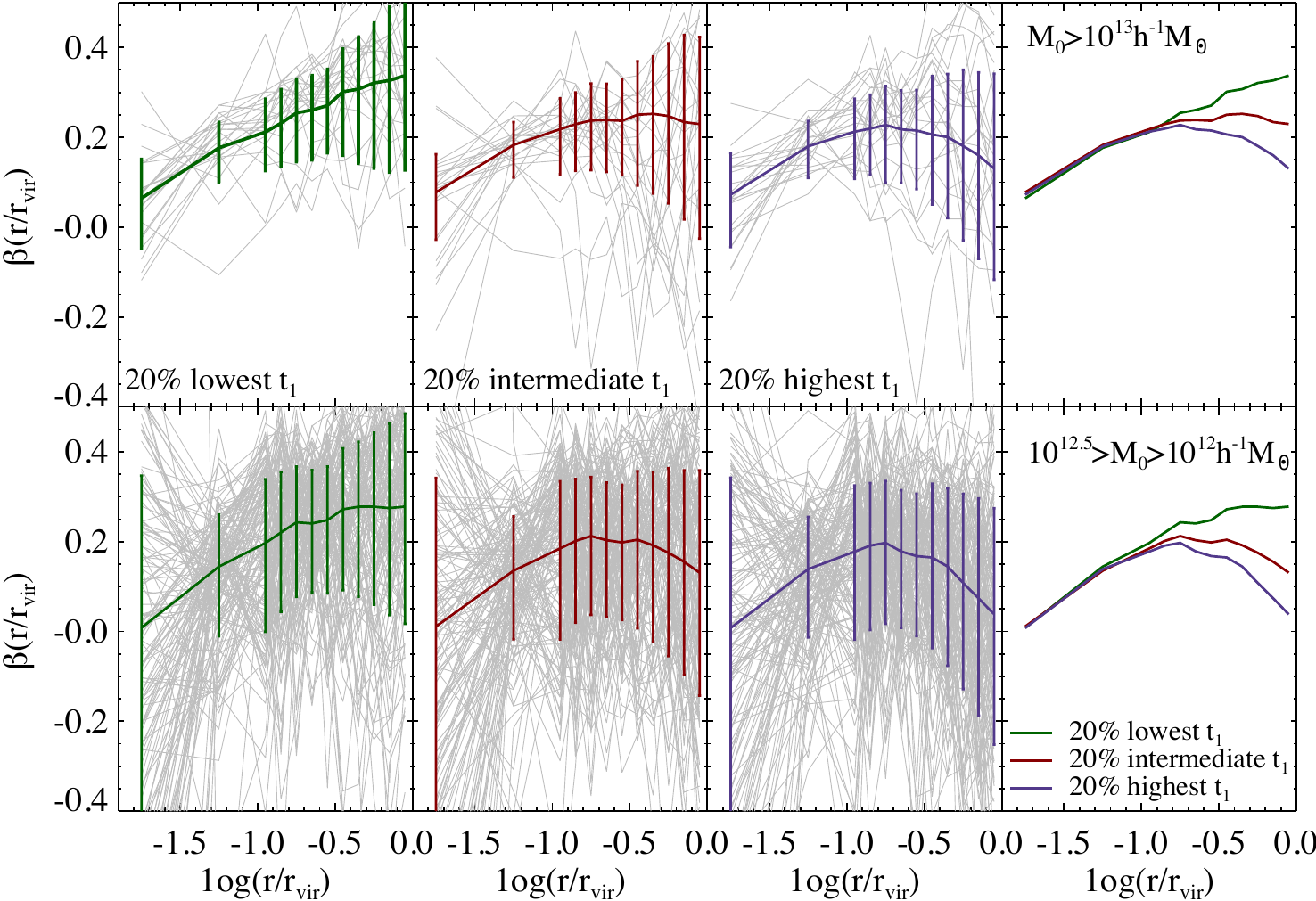,scale=0.8}
\caption{The velocity anisotropy profiles (gray lines) for $z=0$ host halos in different tidal fields,
as indicated in the panels. For clarity, profiles of 2\% of the halos in
each sample are plotted. The median profiles are plotted in colored lines,
with error bars representing the $1\sigma$ scatter.
For comparison, the median profiles for different $t_1$ are re-plotted
in the rightest-hand column.
}
\label{fig_anp}
\end{figure*}

To investigate the velocity anisotropy in more detail,
we estimate the velocity anisotropy profile, $\beta(r/r_{vir})$,
for individual host halos. Here $r$ is the distance to the minimum
potential position in the host halo, and $r_{\rm vir}$ is its
virial radius. Here we use all particles in each radius bin
to calculate the dispersion. Some non-halo particles may be included,
but our test showed the effect is small.
Figure \ref{fig_anp} presents the results separately
for halos residing in the highest, intermediate and lowest 20\% $t_1$
environments. For clarity, in each case, we only show the profiles of
2\% halos randomly selected from the total sample.
The median $\beta$ profiles and the one sigma scatter
are also plotted for reference. The large scatter in
the innermost bins for the low mass halos are due to small
number statistics. These profiles are in broad agreement with
those obtained by \citet{Ludlow_etal2011}
from a much higher resolution simulation.
As one can see from the right panels, the anisotropy profile
depends significantly on the tidal field.
For halos in the lowest 20 percentile of $t_1$ distribution,
the median $\beta$ increases monotonously with increasing radius, indicating
that the orbits of dark matter particles become increasingly
radial as $r$ increases. In contrast, for halos in the highest 20
percentile of $t_1$, the median $\beta$ first increases
and then decreases with $r$, reaching a maximum value
of $\beta\sim0.2$ at $r\sim0.16r_{\rm vir}$. The velocity
dispersion on average approaches isotropy ($\beta\to 0$)
in the outermost regions of such halos. The environmental effect
decreases with decreasing radius, becoming unimportant at
$r< 0.1 r_{vir}$. In the innermost region, the velocity dispersion
is quite isotropic (i.e. $\beta\sim0$), independent of the tidal field.

As shown in \citet{Zhao_etal2003}, a cold dark matter halo
grows in an inside-out fashion after its potential well is established.
Thus, the outer parts of halos are expected to be dominated by
newly accreted material, and the material in the outer part should
contain more information about the recent accretion events.
This is the primary reason why the velocity structure in the outer parts
of halos depends strongly on $t_1$. The behavior in the inner
parts is more difficult to understand. As shown in Figure \ref{fig_t1rv},
radial accretion is actually more dominating at higher redshift.
Since the inner parts are expected to have formed earlier,
the weak anisotropy seen in the inner region cannot be
due to the initial orbits of infall halos. It is possible that non-linear
evolution, such as radial orbit instability
\citep{CarpinteroMuzzio1995, MacMillan_etal2006, Bellovary_etal2008}
have suppressed the initial velocity anisotropy. It is also
possible that the early assembly of halos is more dominated
by major mergers, and the associated violent relaxation
reduces the initial anisotropy \citep{Lu_etal2006}.

Next we consider another important halo property, namely the
angular momentum. Following common practice, we use the
spin parameter,
\begin{equation}\label{lambda_def}
\lambda=\frac{J|E|^{1/2}}{GM_0^{5/2}}
\end{equation}
to characterize the angular moment of a halo,
where $J$ is the angular momentum, $E$ the total energy and
$G$ the gravitational constant. We adopt the method
presented in \citet{Bett_etal2007} to estimate the total energy.
The direction of the angular momentum (the spin direction)
of a halo is denoted by ${\bm j}$.

\begin{figure*}
\centering
\epsfig{file=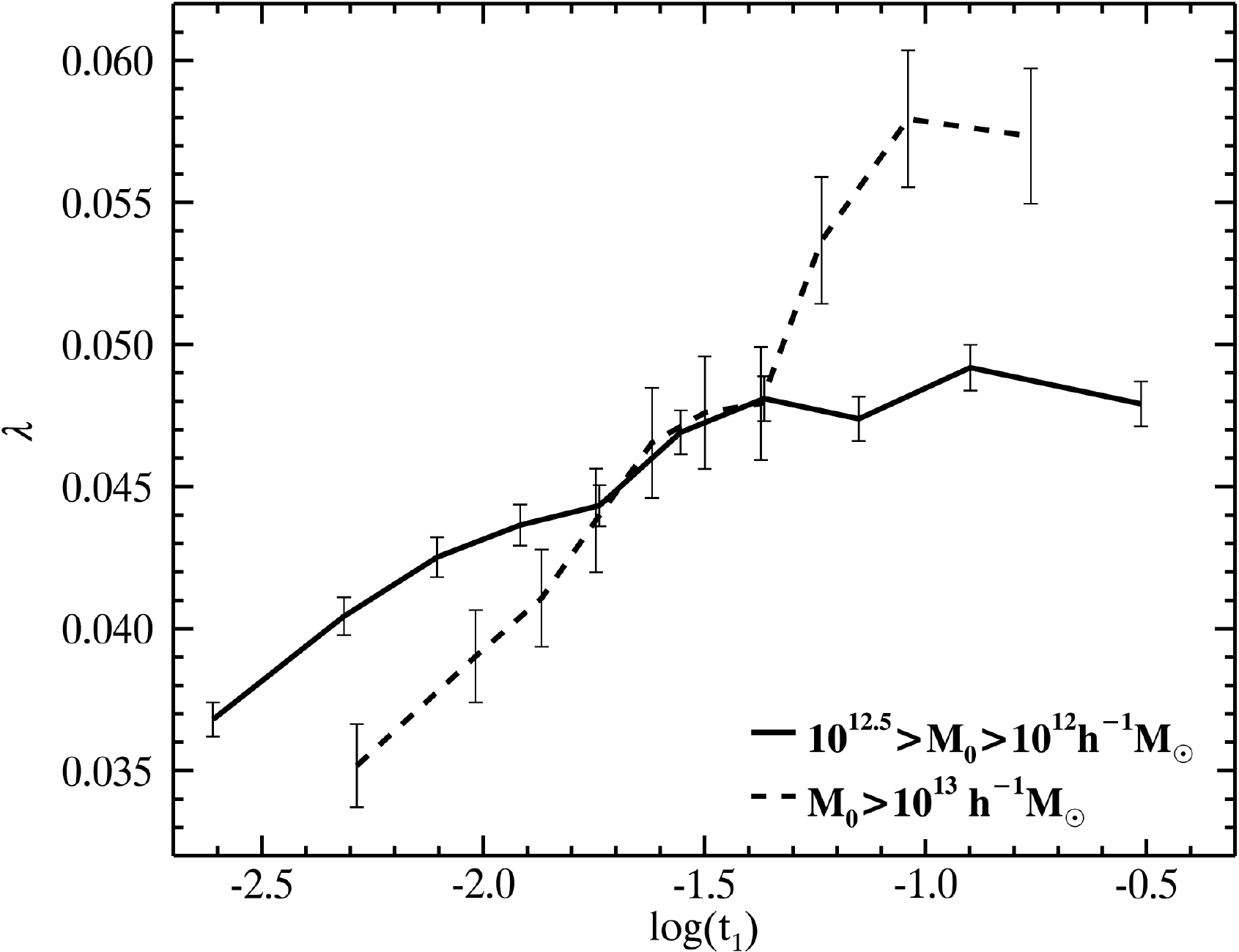,scale=0.8}
\caption{The spin parameter, $\lambda$ defined by
equation (\ref{lambda_def}),
of host halos as a function of $t_1$. All $t_1$
bins contain equal number of halos.
The curves connect the median values of
 $\lambda$, while the error bars are Poisson errors.}
\label{fig_t1lamb}
\end{figure*}

Figure \ref{fig_t1lamb} shows the median $\lambda$ as a function
of $t_1$. Clearly, on average halos spin faster in a stronger tidal field,
and the dependence is stronger for more massive halos.
This result has already been obtained in \citet{Wang_etal2011}
and is consistent with the spin-dependent halo clustering found
by \citet{Bett_etal2007}.

\begin{figure*}
\centering
\epsfig{file=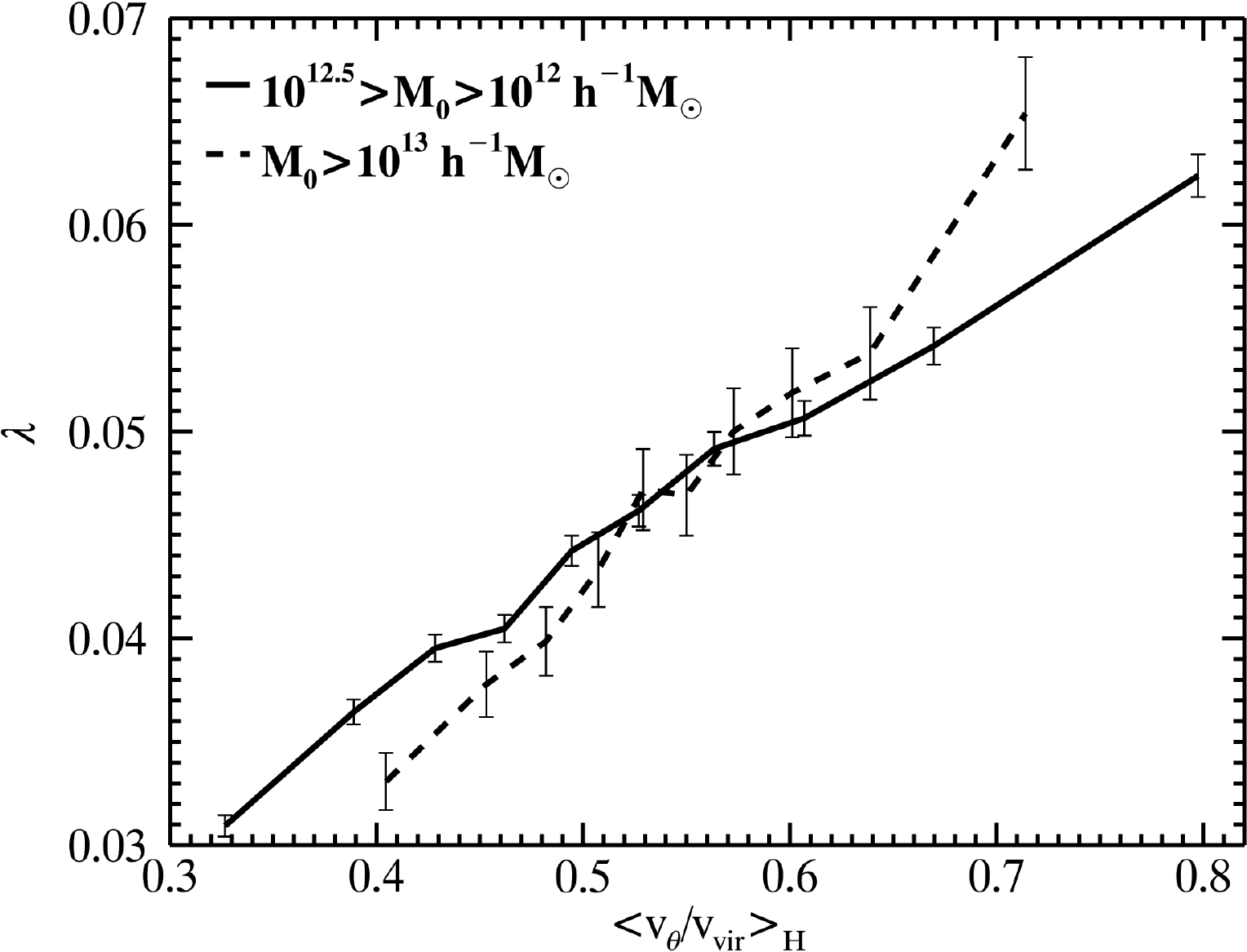,scale=0.8}
\caption{The spin parameter, $\lambda$, as a function of
$\langle v_{\theta}/v_{\rm vir}\rangle_{\rm H}$.
For each host halo, $\langle v_{\theta}/v_{\rm vir}\rangle_{\rm H}$ is
the mass-weighted average value over all of its infall halos in
the staying population. All $\langle v_{\theta}/v_{\rm vir}\rangle_{\rm H}$
bins contain equal number of halos.
The curves connect the median values of
 $\lambda$, while the error bars are Poisson errors.}
\label{fig_lambvt}
\end{figure*}

As shown in Figure \ref{fig_vt}, the tidal field can significantly
enhance the tangential velocities of infall halos, which may in turn
increase the orbital angular momenta of the host halo.  To demonstrate
this, we estimate the mean tangential velocity,
$\langle v_{\theta}/v_{\rm vir}\rangle_{\rm H}$, of infall halos
(the staying population only) for each $z=0$ host. Figure \ref{fig_lambvt}
shows how  $\lambda$ depends on $\langle v_{\theta}/v_{\rm vir}\rangle_{\rm H}$.
As expected, the spin parameter has a strong positive
correlation with the mean tangential velocity of infall halos.
This suggests that halos acquire their angular momenta via the
large scale tidal field which regulate the orbital angular momenta
of infall halos.

\begin{figure*}
\centering
\epsfig{file=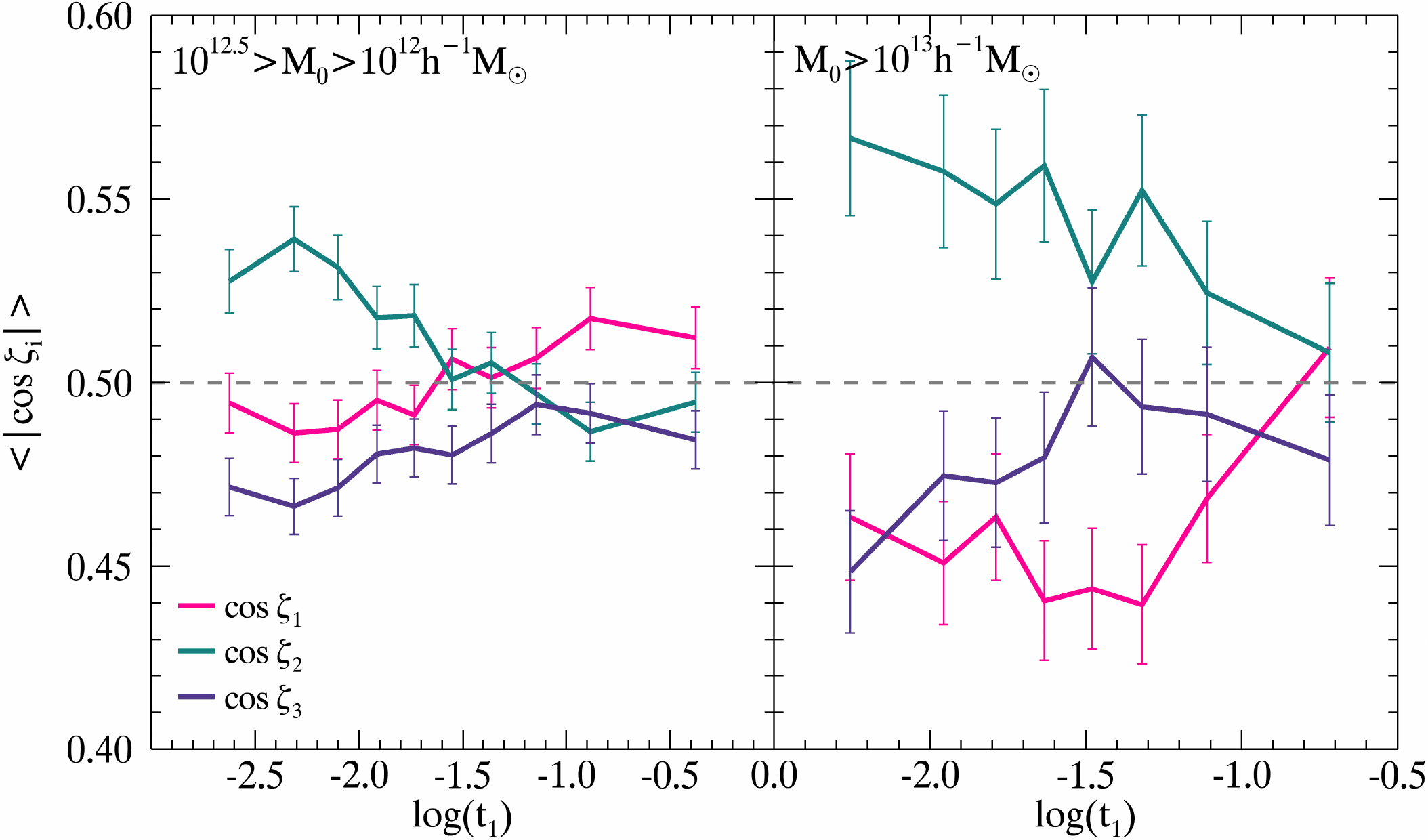, scale=0.7,angle=270}
\caption{The mean of $\vert\cos \zeta_{i}\vert$, a measure of the
alignment between the tidal field and halo spin defined by
equation (\ref{t1_tj}), as a function of tidal field strength $t_1$.
All $t_1$ bins contain equal number of halos.
The error bars are Poisson errors. The horizontal lines
indicate no alignment.}
\label{fig_t1tij}
\end{figure*}

In the literature, halo angular momenta are believed to be generated
by tidal torques of the large scale structure  \citep[e.g.][]{Porciani_etal2002}.
One unique prediction of the tidal torque theory is that the halo spin
axis tends to be parallel with the intermediate axis the tidal field, i.e. ${\bm t}_2$,
and perpendicular to ${\bm t}_1$ and ${\bm t}_3$. \citet{Wang_etal2011}
detected such alignments in their simulations, but the signals are
rather weak \citep[see also][]{Forero_Romero_etal2014}. 
Some studies also found that halo spin
tends to be perpendicular to filament and parallel to sheet 
\citep[e.g.][]{Hahn_etal2007, Zhang_etal2009, Libeskind_etal2013apj}, 
which is consistent with the alignment with the intermediate axis 
of the tidal field.
As we have already demonstrated in Section \ref{sec_eha},
accretion patterns are quite different between strong and weak tidal fields,
and so the alignment signals may also vary with the strength of the local tidal
field. To test this, we study the angle $\zeta_{i}$ between ${\bm j}$ and
${\bm t}_{i}$:
\begin{equation}\label{t1_tj}
\cos (\zeta_i) = {{\bm j} \cdot {\bm t}_i \over
\vert {\bm j}\vert \vert {\bm t}_i\vert}\,
~~~(i=1,2,3)\,.
\end{equation}
Figure \ref{fig_t1tij} shows the mean of $\vert \cos\zeta_i\vert$ as a
function of $t_1$. As one can see, when the tidal field is weak, the alignments
are perfectly consistent with the prediction of the tidal torque theory,
in that ${\bm j}$ tends to align with ${\bm t}_2$.
However, as the tidal field gets stronger, the alignments become weaker and weaker.
For small halos in regions  of high $t_1$, the trend is  eventually reversed
so that  the spin tends to be aligned with ${\bm t}_1$ and perpendicular to ${\bm t}_2$.
This reversal is caused by the strengthened tidal truncation
of accretion along the ${\bm t}_1$ direction where the tearing by the tidal field
is the strongest. In an analysis of spin alignments using all halos (or galaxies)
without regarding their local tidal fields, the opposite trends in
strong and weak fields may cancel each other and weaken the total
signal. It is thus important to take into account the local tidal field
strength when investigating spin alignments in both observation and
numerical simulation.

Finally, we examine the second moment tensor of the internal
velocity field of a halo, defined as
\begin{equation}
I^v_{jk}=\sum_{n} v_{n,j}v_{n,k}\,,
\end{equation}
where $v_{n,j} (j=1,2,3)$ are the three velocity components of
the $n$th particle in the halo. The
square root of the eigenvalues of the tensor can be
used to represent the principal axes, $I^v_1$, $I^v_2$ and $I^v_3$
($I^v_1\geq I^v_2\geq I^v_3$), and the axis ratios,
such as $I^v_3/I^v_1$,  to characterize the velocity ellipsoid.
The corresponding eigenvectors, ${\bm I}^v_1$, ${\bm I}^v_2$ and ${\bm I}^v_3$,
represent the directions of the major, intermediate and minor axes of the
velocity ellipsoid, respectively.

\begin{figure*}
\centering
\epsfig{file=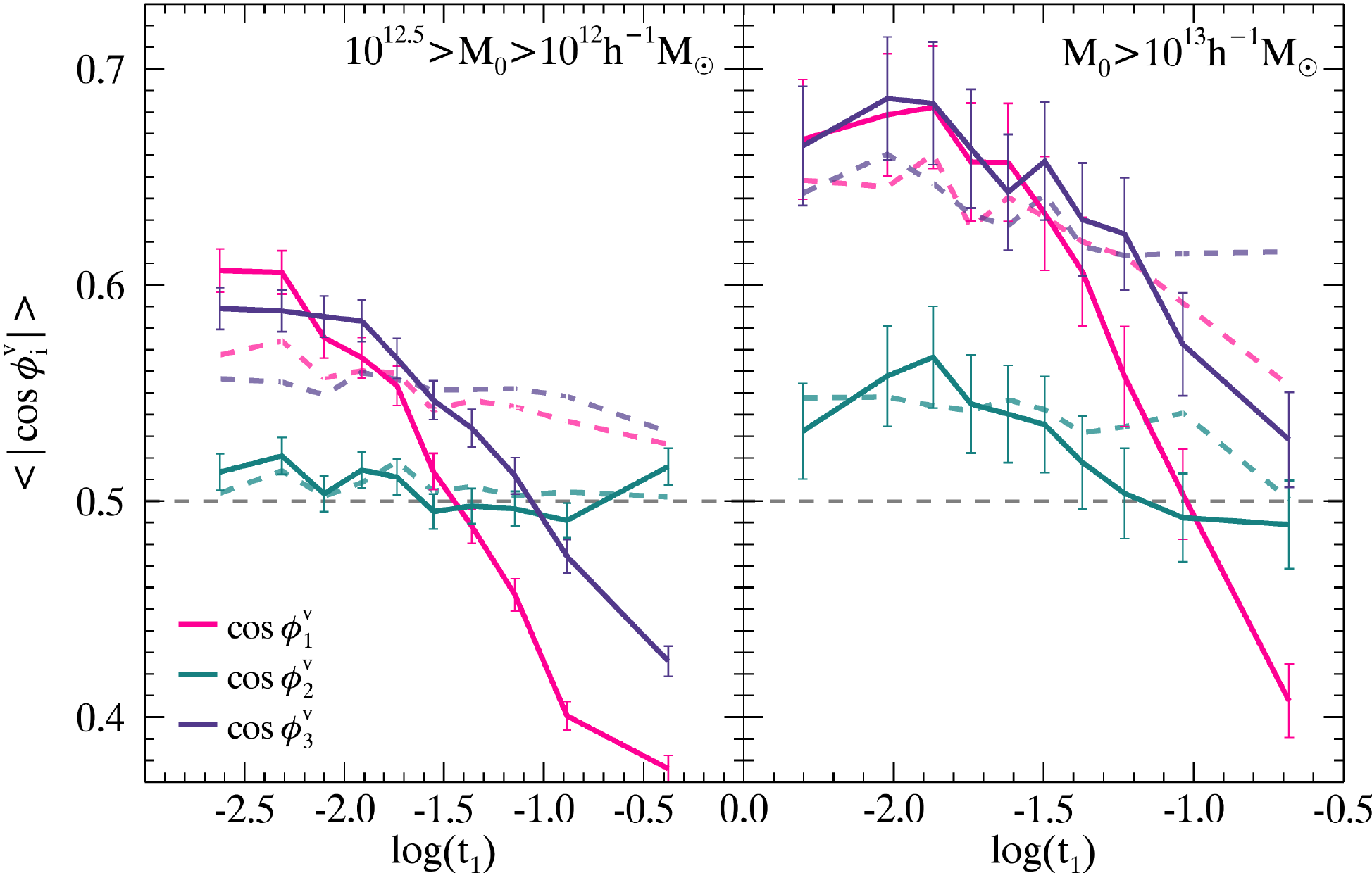,scale=0.7,angle=270}
\caption{The mean of $\vert\cos \phi_{i}^{v}\vert$,
defined by equation (\ref{phi_It}),
as a function of $t_1$. All $t_1$ bins contain equal number of halos.
The solid lines are obtained from using all particles in a halo,
while the dashed lines use halo particles in the inner part
with $r\le 0.1r_{vir}$.  The Poisson errors, which are the same
for both the solid and dashed curves at a given $t_1$, are
shown only on the solid curves.
The horizontal lines indicate isotropic distribution.}
\label{fig_t1ii}
\end{figure*}

To check the alignment between the eigenvectors of the velocity
tensor and the tidal field, we use the angle $\phi^v_i$ defined as
\begin{equation}\label{phi_It}
\cos (\phi^v_i) = {{\bm I}^v_i \cdot {\bm t}_i \over
\vert {\bm I}^v_i\vert \vert {\bm t}_i\vert}\,
~~~(i=1,2,3)\,.
\end{equation}
Figure \ref{fig_t1ii} shows the mean of $\vert \cos\phi^v_i\vert$
as a function $t_1$. In weak tidal field, ${\bm I}^v_1$ tends to be parallel
with ${\bm t}_1$, but an opposite trend is seen in strong tidal field.
The trend in $\vert\cos\phi^v_3\vert$ is very similar but slightly weaker,
and there is no significant alignments between  ${\bm I}^v_2$ and
${\bm t}_2$ regardless of tidal field strength. The transition is similar
to what is seen in the relationship between $\cos\theta^v_i$ and $t_1$
(see Section \ref{sec_ali}), and it may be possible that the results
shown in Figure \ref{fig_t1ii} can be understood in terms of halo accretion.
As shown in Figure \ref{fig_t1vt}, the velocities of infall halos
tend to be parallel with ${\bm t}_1$ and perpendicular to ${\bm t}_3$
in weak tidal field. If the host halo retains the velocity structure
of the infall halos, the major (minor) principal axes of its velocity ellipsoid
are expected to be parallel with ${\bm t}_1$ (${\bm t}_3$), as shown
in Figure \ref{fig_t1ii}.  Similarly, the different accretion pattern at
high $t_1$ can also explain why ${\bm I}^v_1$ (${\bm I}^v_3$) tends
to be perpendicular to ${\bm t}_1$ (${\bm t}_3$) in strong tidal fields.

\begin{figure*}
\centering
\epsfig{file=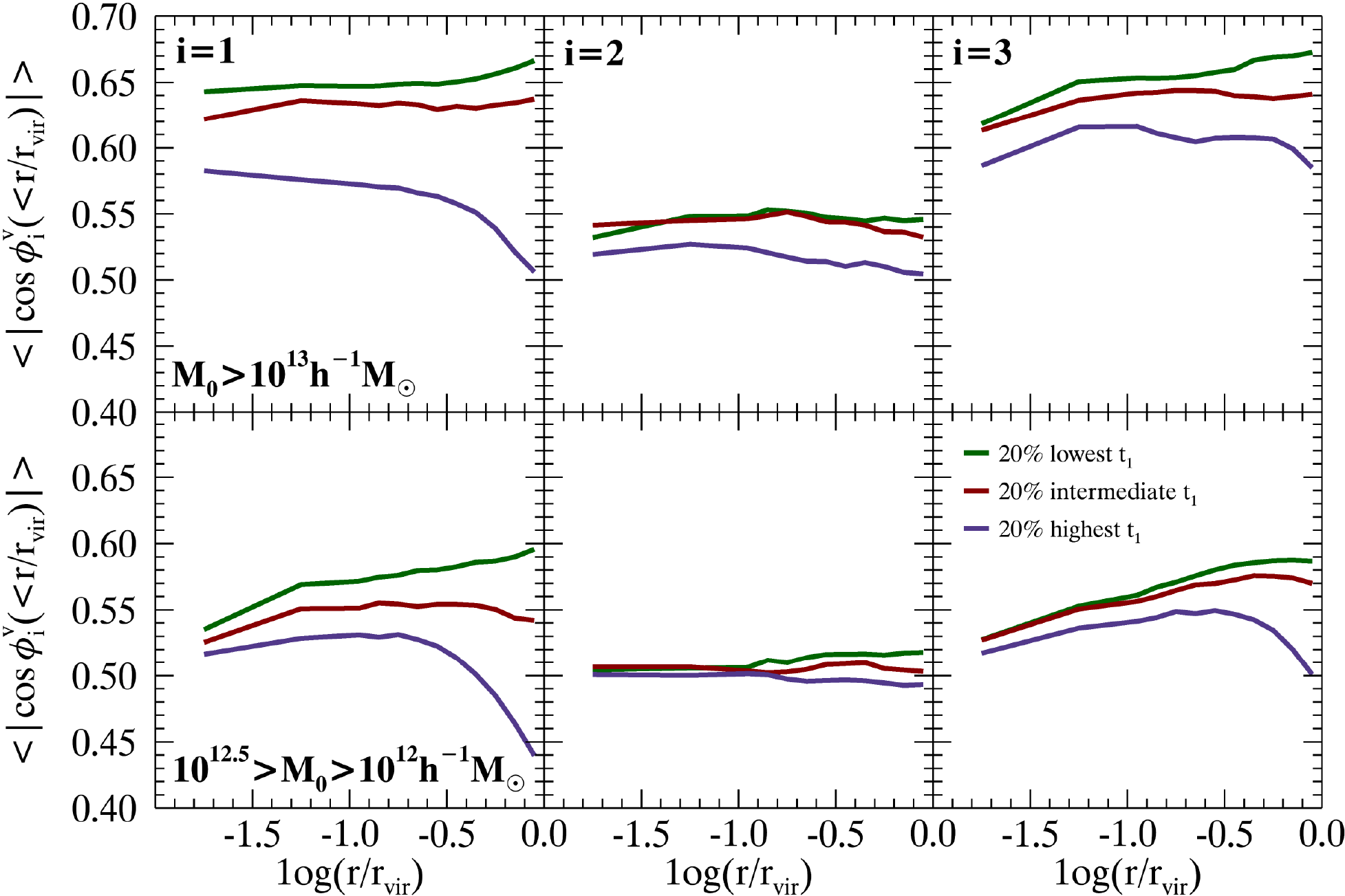, scale=0.7,angle=270}
\caption{The mean alignment profile within individual halos
located in tidal fields of three different $t_1$ 20 percentile intervals.
Here $\vert \cos\phi^v_i(<r/r_{vir})\vert$ measures
the alignment between tidal field tensor and velocity ellipsoids
obtained using particles within $r/r_{vir}$.}
\label{fig_veprof}
\end{figure*}

In is interesting to see how the velocity ellipsoid in the inner halo
regions, where galaxies are located,  are aligned with the local
tidal field. To do this, we calculate the alignment between the tidal field
and the velocity ellipsoid of the halo particles within some radius
$r$. The dashed curves in Figure \ref{fig_t1ii} show the results
for $r=0.1r_{\rm vir}$, and the averages of $\vert \cos\phi^v_i(<r/r_{vir})\vert$
are shown as functions of $r/r_{vir}$ in Figure \ref{fig_veprof}.
It is evident that the velocity ellipsoids of the inner halo regions align
with the tidal field in a different way from the whole halos.
In weak tidal fields, the alignment signals in the inner regions are
slightly weaker than those for the whole halos. In strong tidal field,
however, the alignments of the inner ellipsoids one average are
opposite to the whole halo. For example, for low mass halos,  the major
and minor principal axes of the inner velocity ellipsoids tend to be parallel with
${\bm t}_1$ and ${\bm t}_3$, respectively, but they tend to be
perpendicular to each other for the whole halos. This suggests
that the accretion pattern at early epoch is similar to that in weak tidal
field environments at low redshift.

\section{Discussion and Summary}
\label{sec_sum}

In this paper, we investigate the environmental dependence of
halo accretion and their impact on the halo dynamical properties. We construct
halo merger trees from $N$-body simulations and identify infall halos that are
about to merge with their hosts. The infall halos are divided into two populations:
the staying population which remain as subhalos within their hosts at $z=0$;
the ejected population which are ejected by their hosts at presented day. We use
the large scale tidal field estimated from the halo population at $z=0$ as
environmental indicator.

We first investigate the infall halo mass functions in various tidal fields.
The mass function for the staying population is quite independent of
both the tidal field and host halo mass. In contrast, the ejected halo mass function
depends strongly on the tidal field. In a stronger tidal field, infall halos are more
easily ejected by their hosts at $z=0$, and smaller infall halos are more
likely to be ejected than massive ones.

We then check the dependence of the orbital parameters of infall halos on
local tidal field. The tidal field does not significantly affect the average radial velocities
of infall halos, but can generate tangential motions. Consequently,
infall halos in stronger tidal fields tend to have higher mean tangential velocities,
larger infall angles and higher velocity dispersions. These results suggest that tidal field tends to pull the accretion flow into 
the orbit of the host and enhance the velocity dispersion among the infall halos, 
making the accretion flow `hotter' and more difficult to capture by the host.

We find that the accretion patterns are different between strong and weak tidal fields.
In weak tidal fields, the positions of infall halos relative to their hosts have a strong
tendency to be parallel with the stretching direction of the tidal field, ${\bm t}_1$, and
perpendicular to the compressing direction, ${\bm t}_3$. Similarly, the velocities
tend to be parallel with ${\bm t}_1$ and  perpendicular to ${\bm t}_3$. The situation
in strong tidal fields is rather different, or even the opposite: the alignments between
the position and tidal vectors become weaker or even absent, and
the velocities tend to be perpendicular to ${\bm t}_1$ and parallel with ${\bm t}_3$.
Such difference is particularly strong for infall halos at low redshift around
low-mass hosts.

The ejected population shows very different behavior from the staying
population. They have much higher tangential velocities. In particular, the velocities
have a strong tendency to be perpendicular to ${\bm t}_1$ and parallel with ${\bm t}_3$
in both weak and strong tidal fields. Ejected halos are more
abundant in stronger tidal fields because the larger and more
tangential velocities generated by the tidal forces make the
infall halos more difficult to hold by their hosts.

The environmental effects of halo accretion are imprinted on the halo
dynamical properties. The environmental dependence of infall angles
results in a correlation between the velocity anisotropy and tidal field.
In particular, the large scale tidal field can affect the anisotropy down to
radius much smaller than the virial radius. Tidal field increases the
tangential velocities of infall halos, producing a positive correlation
between halo spin and the strength of local tidal field.
The alignment signal of spin with tidal field is different between weak
and strong tidal fields; only in weak tidal fields does
the spin-tidal field alignment follow the prediction of the tidal
torque theory, with the spin parallel to the intermediate axis
and perpendicular to the major axis of the tidal tensor,
while in strong fields the alignment is the opposite, at least
for low-mass halos.   Finally, we find a dramatic transition in the
alignment between the principal axes of halo velocity ellipsoid
and the tidal field tensor in strong and weak tidal field,
A radial dependence for this alignment is also found,
which differs between weak and strong tidal fields.
All these indicate that large scale tidal field affects halo
dynamical properties via regulating the flow patterns
around halos.

Our results suggest the tidal field describes well the following
two aspects of environmental effects. First, strong tidal field
(i.e. larger $t_1$) tends to make the surrounding environment
`hotter', thereby boosting the fraction of the ejected
sub-halos, and increasing the tangential component of velocity
and the velocity dispersions. This affects halo dynamical properties,
making the internal velocity field less radial and boosting
the angular momentum in a way that is different from the predictions
of the tidal torque theory. Second, the local tidal field describes
well the local density and velocity structures from which
material is accreted in halos, which in turn explains how halo
intrinsic properties are correlated with local tidal fields.

Our results have important implications for observations. For example,
our results suggest that the correlation between spin axes of
disk galaxies and the large scale structure should be studied separately
for the weak and strong tidal fields. The fact that strong alignments
of halo velocity ellipsoids with local tidal fields extend all the way
to halo central parts suggests that the orientations of elliptical
galaxies should be tightly correlated with the local tidal fields. The
predicted dependence of such alignments on the strength of local
tidal fields can also be tested using a large sample of elliptical galaxies.
Furthermore, the infall patterns around halos can be studied
by using the distributions and velocity fields traced by satellite
galaxies in and around  dark matter halos represented
by galaxy groups.  Since the tidal fields enhance
the tangential velocities of infall halos, we may expect the
dynamical time scales relevant to mergers to be larger for
those located in stronger tidal fields. This difference in
dynamical time scales may lead to differences in the abundance
and properties of the satellite population.  We will come back to
some of these problems in our future work.

\section*{Acknowledgments}

We thank Volker Springel for kindly providing his SUBFIND code.
The numerical calculations have been done on the supercomputing system in the Supercomputing Center of University of Science and Technology of China. This work is supported by the Strategic Priority Research Program "The Emergence of Cosmological Structures" of the Chinese Academy of Sciences, Grant No. XDB09010400, NSFC (11421303), 973 program (2015CB857005), the Fundamental Research Funds for the Central Universities, HJM would like to acknowledge the support of NSF AST-1109354.

\newpage

\end{document}